\documentclass[times,authoryear,3p,twocolumn]{elsarticle}

\usepackage{natbib}
\usepackage{txfonts}
\usepackage{graphicx}
\usepackage{rotating}

\newcommand{\gro}{GRO~J1655--40} 
\newcommand{\grs}{GRS~1915+105} 
\newcommand{\aaa}{A0620--00} 
 
\newcommand{\avx}{$A_V^{\textrm{\scriptsize X-rays}}$} 
\newcommand{\avo}{$A_V^{\textrm{\scriptsize optical}}$} 
\newcommand{\kms}{~km~s$^{-1}$}

\begin{document}

\title{What is the closest black hole to the Sun?\tnoteref{t1}}
\tnotetext[t1]{Based on data obtained at the European Southern Observatory (La Silla and Paranal Observatories), under program IDs 276.D-5027(A), 066.D-0157(A), 70.D-0766(A)  and 074.D-0204(A).} 

\author[cf]{C\'edric Foellmi\corref{cor1}}
\ead{cedric.foellmi@obs.ujf-grenoble.fr}
\cortext[cor1]{Corresponding author} 
\address[cf]{Laboratoire d'Astrophysique de Grenoble, Observatoire de Grenoble, Universit\'e Joseph Fourier, CNRS UMR 5571, 414 Rue de la Piscine, 38400 Saint-Martin d'H\`eres, France} 

\begin{abstract}
We examine the distance of the two galactic microquasars \gro\ and \aaa\, which are potentially the two closest black holes to the Sun. We aim to provide a picture as wide and complete as possible of the problem of measuring the distance of microquasars in our Galaxy. The purpose of this work is to fairly and critically review in great detail every distance method used for these two microquasars in order to show that the distances of probably all microquasars in our galaxy are much more uncertain than currently admitted. Moreover, we show that many confirmations of quantitative results are often entangled and rely on very uncertain measurements.  We also present a new determination of the maximum distance of \gro\ using red clump giant stars, and show that it confirms our earlier result of a distance less than 2~kpc  instead of 3.2~kpc. Since it then becomes more likely that \gro\ could originate from the stellar cluster NGC~6242, located at 1.0 kpc, we review the distance estimations of \aaa, which is so far the closest black hole with an average distance of about 1.0~kpc. We show that the distance methods used for \aaa\ are also problematic. Finally, we present a new analysis of spectroscopic and astrometric archival data on this microquasar, and apply the maximum-distance method of Foellmi et al. (2006). It appears that \aaa\ could indeed be even closer to the Sun than currently estimated, and consequently would be the closest known black hole to the Sun.
\end{abstract}


\maketitle

\section{Introduction}

Microquasars are stellar binaries in which one of the component is a black hole. The stellar companion is filling its Roche lobe, ejecting matter through the first Lagrangian point, which then organizes itself around the compact object as an accretion disk. The temperature of the disk increases strongly towards the center and large amounts of X-rays with powerful, persistent and collimated jets are produced. In some cases, the relativistic jets appear to be superluminal. Microquasars are galactic laboratories of high-energy phenomena, and they must be seen as part of a large paradigm where AGNs, microquasars and possibly gamma-ray bursts (GRBs) share similar physics \citep{Mirabel-2004}. 

There are various physical reasons why the distance is an important parameter in our understanding of microquasars. Firstly, the superluminal velocity effect of the jets \citep[which is a relativistic time delay effect producing an apparent velocity on the sky larger than the speed of light; see e.g.][]{Rees-1966} depends on the distance. Although it is a pure geometrical effect, it has a minimum threshold of $\beta = v/c \geq 2^{-1/2}$ (with $c$ the speed of light) below which the jets are not superluminal, whatever their true spatial speed $v$. When one tries to model the jets, and more importantly, the jet's launching mechanism, it is important to know the true velocity (and hence power) of the jets. Moreover, this effect is not very common, and it will become more important to correctly identify superluminal objects when the statistics will grow with future X-ray missions.

Secondly, the possible misalignment of the jets with respect to the accretion disk axis. As a matter of fact, \gro\ is often quoted as the typical microquasar where the jets and the disk are misaligned \citep[e.g.][]{Maccarone-2002}. This is directly related to the fact that the determination of the jet projection angle is linked to that of the distance, as explained below. The possible misalignment of the jet is an important clue of the formation and evolution of the system \citep[see e.g.][]{Martin-etal-2008}.

Thirdly, the black holes' origin, and their orbit in our Galaxy. This is particularly true for \gro, where a change in the distance by a factor of three makes a large difference to the orbit of the object in the Galaxy. See for instance \citet{Mirabel-etal-2002} who have calculated the orbit for \gro\ with $D = 0.9$ and 3.2 kpc. This is important for our understanding of the origin of black holes in our Galaxy, and their distribution throughout the disk. 

Finally, the true luminosity of the stellar companion can also be an important point in our understanding of the microquasars as dynamical objects, and in particular how the companion is affected by the filling of its Roche lobe and is certainly (partially) irradiated by X-rays from the accretion disk \citep[e.g.][]{Dubus-etal-1999}. 

The distance of microquasars is a simple yet central parameter in our description of these objects, conditioning a large part of our understanding both of the physics and the astronomical views of black hole stellar systems. 

\subsection{The origin of the disagreement on the distance of \gro}

The origin of this work is to be found in the publication by \citet{Mirabel-etal-2002} who note that the distance of \gro\ can be actually radically different from the accepted distance of 3.2~kpc determined by \citet{Hjellming-Rupen-1995}. This later value has since then been apparently confirmed by many other studies. In fact, \gro\ is a runaway black hole and Mirabel et al. published its proper motion, obtained with the {\it Hubble Space Telescope}. The opposite direction of the proper motion points almost perfectly (see their Fig. 1) to the center of a cluster (NGC~6242) located at 1.0 kpc from the Sun \citep{Kharchenko-etal-2005}. It is then tempting to think that the system received a kick velocity at the moment of the primary's supernova explosion, and then moved away from its original cluster. This is the starting question: is the distance of \gro\ 3.2 or 1.0 kpc? Among other things, the jets are superluminal with $D=3.2$ kpc, but not at 1.0 kpc.

\citet{Foellmi-etal-2006b} published a new method providing only an upper limit to the distance and giving $D\lesssim 1.7$ kpc when applied to \gro, thus challenging the accepted distance of 3.2~kpc. This method is also partially problematic, and it will be discussed critically below. With a maximum distance of 1.7~kpc, \gro\ becomes a likely candidate of being the closest (stellar) black hole to the Sun. The position is currently held by \aaa, which has an average measured distance of about 1.0 kpc. We will show however that the distance of \aaa\ is also very uncertain, revealing different problems in each distance method.

In summary, none of the published distances of \gro\ and \aaa\ are so far reliable. This paper emphasizes the difficulties encountered when determining the distance of microquasars, and shows that the distances of probably many, if not all, microquasars in our Galaxy are much more uncertain than currently admitted. In this paper we aim to provide a fair and critical review on the distance methods used on these two galactic microquasars. Other microquasars will be the subject of a subsequent work. In addition to \citet{Foellmi-etal-2006b}, partial material (incomplete and partially incorrect on the radio-jet distance of \gro) has already been published in \citet{Foellmi-2006e} and \citet{Foellmi-2007c}. This paper addresses many more important details, and in particular the issues of the dynamical studies used to determine the absolute magnitude of the secondary star of \gro. Moreover, we also provide a completely new estimate of the distance of \gro. Finally, we apply the method of \citet{Foellmi-etal-2006b} to \aaa, along with archival astrometric data. We conclude that \gro\ is certainly closer than the current accepted distance, but that \aaa\ might be even closer still.

\subsection{Distance methods: the basics}

The distance of a {\it stellar} object is often measured by comparing its absolute and apparent magnitudes. There are other methods, such as astrometry and parallax, and jet speed measurements (see below) for the special case of microquasars. But most often, the core method is simply this one. It reads: 
\begin{equation}
	m_{\textrm{\scriptsize true}} - M = 5 \, \log(D) - 5 
\end{equation}
where $m_{\textrm{\scriptsize true}}$ is the true apparent magnitude of the object, affected only by the geometrical distance separating the object and the observer.

There are (at least) two major issues with this method: one is general, and one is specific to stellar binaries of short period. The first {\it observational} difficulty is that $m_{\textrm{\scriptsize true}}$ is usually different from the observed apparent magnitude $m_{\textrm{\scriptsize obs}}$ since the light is going through some absorbing patchy interstellar medium, which makes the star dimmer and redder. We have: $m_{\textrm{\scriptsize true}} = m_{\textrm{\scriptsize obs}} - A$, where $A$ is the absorption, in the given passband, expressed in magnitudes. Therefore: 
\begin{equation}
	m_{\textrm{\scriptsize obs}} - M - A = 5 \, \log(D) - 5 
	\label{equ_magdist} 
\end{equation}
The determination of $A$ is critical. It is often not measured directly, but rather is the so-called color excess, or reddening: $E(B-V)$, which is the relative amount of additional red color between the $B$ and $V$ bands due to the absorption of bluer wavelengths by the gas and the dust. For a star, the color excess can be obtained by comparing the observed and the intrinsic $B$ and $V$ color indices (the latter being inferred from the spectral type): 
\begin{equation}
	E(B-V) = (B-V)_0 - (B-V) 
	\label{equ_ebmv}	
\end{equation}
The absorption in the $V$ band follows: 
\begin{equation}
	A_V = R \times E(B-V) 
	\label{equ_AV} 
\end{equation}
where $R$ is the total-to-selective absorption. For other colors, this relation needs corrections \citep[see e.g.][]{Fitzpatrick-1999}. The practice shows that what is often measured directly is $E(B-V)$, $R$ being often approximated\footnote{The difference between $R=$3.0 and 3.7 implies a relative distance uncertainty of 1.38.} by a value between 3.0 and 3.7 with a more-or-less canonical value of 3.1. In fact, $R$ is a function of the interstellar reddening curve and the color of the stars, because the wide passbands of the photometric $B$ and $V$ filters makes the effective wavelengths of the filters shift with different stellar intensity gradients. \citet{Olson-1975} gives an approximate relation for $R$: 
\begin{equation}
	R = 3.25 + 0.25\, (B-V)_0 + 0.05\, E(B-V) 
	\label{equ_R}	
\end{equation}
where $(B-V)_0$ is the unreddened color index of the star. This relation is valid for normal stars with $(B-V)_0 < 0.4$ and $E(B-V) < 1.5$ to within an error of 0.05 in $R$ \citep{Olson-1975}. See for instance \citet{Crawford-Mandwewala-1976} for a comparison for various photometric systems, and \citet{McCall-2004} for an updated discussion on this topic.

The other issue with the method consisting of comparing the magnitudes, specific to the stellar binaries with short period, is that the stellar companion (here the secondary star) is certainly not spherical anymore, since it completely fills its Roche lobe to feed the accretion disk. Moreover, its surface temperature might also not  be homogeneous because of irradiation \citep[e.g.][]{Dubus-etal-1999,OBrien-etal-2002}, making its average observed temperature a function of the orbital phase. Hence uncertainties arise when one tries to estimate the true value of the absolute magnitude, $M$, in equation~\ref{equ_magdist}, since it is often calibrated with "normal" (spherical) and isolated stars. 

The absolute magnitude is normally obtained through the determination of the spectral type of the star, which gives the temperature $T_{\textrm{{\scriptsize eff}}}$, and the modelling of either the radial-velocity curve or the multi-color lightcurves of the binary system gives the size of its orbit. From the latter, one can compute the {\it effective} Roche-lobe radius \citep[see for instance][]{Eggleton-1983}, which is then identified to the radius of the star. Assuming a uniform temperature distribution across the star and that it is roughly spherical, the luminosity (and thus the absolute magnitude) can be estimated with:
\begin{equation}
	L = 4 \pi \sigma R^2 T^4
	\label{equ_lum}
\end{equation}
where $\sigma$ is the Stefan-Boltzmann constant.

\section{The published distance of GRO J1655-40}

\gro\ (a.k.a. Nova Sco 94) has been discovered as a Soft X-ray Transient (SXT) on July 27, 1994 with \emph{BATSE} on board the \emph{Compton Gamma Ray Observatory} \citep{Zhang-etal-1994}. Its jets were observed in the radio, giving a distance of about 3 kpc, implying that they are superluminal: 1.5$\pm$0.4~$c$ \citep{Tingay-etal-1995}, 1.05~$c$ \citep{Hjellming-Rupen-1995}, where $c$ is the speed of light. \gro\ was the second superluminal source in our Galaxy shortly after the discovery of the jets in \grs\ by \citet{Mirabel-Rodriguez-1994}. The value of 3.2 kpc has been published by \citet{Hjellming-Rupen-1995} based on VLA and VLBA radio data.

Since then, many publications use directly this canonical distance. For instance: \citet{Brandt-etal-1995,Barret-etal-1996,Regos-etal-1998,vanderHooft-etal-1998,Phillips-etal-1999,Shahbaz-etal-1999,Kuulkers-etal-2000,Soria-etal-2000,Greene-etal-2001,Combi-etal-2001,Buxton-Vennes-2001,Yamaoka-etal-2001,Gierlinski-etal-2001,Kubota-etal-2001,OBrien-etal-2002,Remillard-etal-2002,Kong-etal-2002,Kobayashi-etal-2003,Stevens-etal-2003,Willems-etal-2005,Brocksopp-etal-2006, Miller-etal-2006,Caballero-Garcia-etal-2007,Martin-etal-2008} and \citet{Chakrabarti-etal-2008}. More recently \citet{Shaposhnikov-Titarchuk-2009} use \gro\ and its parameters as a reference object for determining the mass of Cyg~X-1. But a significant number of studies seem to confirm the canonical distance. We challenge here not only the lower limit of 3.0~kpc of \gro\ but also every published confirmation.

We note however that \citet{Migliari-etal-2007} use with caution the upper limit of 1.7~kpc determined in \citet{Foellmi-etal-2006b}.

\subsection{The radio-jet kinematic distance of 3.2~kpc} \label{radio-distance}

\citet{Hjellming-Rupen-1995} present new radio data from which they infer a value for the distance of \gro. The method is simple: the opposite motions of the receding and approaching jets ($\mu_-$ and $\mu_+$ respectively) are directly related to jet projection angle relative to the line of sight $\theta$, the true jet speed $\beta=v/c$, and the distance $D$. The authors measure the intrinsic proper motions of each jet: 54 mas~d$^{-1}$ for the NE component, and 45 mas~d$^{-1}$ for the SW component, and use the kinematic equation described in \citet{Mirabel-Rodriguez-1994}: 
\begin{equation}
	\label{mumu1} 
	\mu_{\pm} = \frac{\beta \sin(\theta)}{1 \pm \beta \cos(\theta)} \frac{c}{D} 
\end{equation}
where $c$ is the speed of light. There are two equations and three unknowns. A constraint on the distance can be obtained by eliminating $\theta$ and requiring that $\beta < 1$: 
\begin{equation}
	\label{mumu2} 
	\frac{D}{c} \left( \frac{2 \mu_+ \mu_-}{\mu_+ + \mu_-} \right) = \frac{v}{c} < 1 
\end{equation}
Taking $v = c$, the maximum distance of \gro\ inferred from the proper motion of the radio jets is $D<3.5$ kpc.

A constraint on the inclination angle of the jets can also be obtained by eliminating $v/D$ and requesting that $v < c$. We obtain: 
\begin{equation}
	\label{mumu3} 
	\frac{v}{c} = \frac{(\mu_- - \mu_+)}{(\mu_- + \mu_+)} \cos(\theta)^{-1} < 1 
\end{equation}
This gives: $\theta \leq 84.8^{\circ}$. Rearranging equation~\ref{mumu1} and eliminating $\beta$, we can write, using the value of $\theta$: 
\begin{equation}
	\label{mumu4} 
	\frac{1}{v/D} \, \left( \frac{2 \mu_+ \mu_-}{\mu_+ + \mu_-} \right) = \sin \theta \leq 0.996 
\end{equation}
that is $v/D >$ 49.3 mas~d$^{-1}$. All these results are given in \citet{Hjellming-Rupen-1995}. 

At this point the authors mention that: "{\it For a distance of 3.2~kpc, this corresponds to $v \geq 0.91 c$, implying $84.3^{\circ} \leq \theta \leq 84.8^{\circ}$.}" Why 3.2 kpc? The only previous mention of the distance at the beginning of the article is only stating that the source lies "at a distance of about 3 kpc", and for which three references were given (all discussed below): \citet[][]{McKay-Kesteven-1994, Harmon-etal-1995, Tingay-etal-1995}. In fact, the value $D = 3.2$ kpc is constrained to lie between the value of 3.0 kpc considered as a lower limit and determined elsewhere, and the actual result on the upper limit of 3.5 kpc. However, choosing 3.2~kpc fixes $\theta$ that is now constrained to a very narrow range ($0.5^{\circ}$), and the system is now defined. 

\subsection{The confirmation of 3.2~kpc with possible wiggles inferred from unaligned images}

Furthermore, this value of 3.2 kpc is being confirmed by a model of the possible jet precession (adding  two parameters to the kinematic model: the precession period and axis inclination). It is however barely relevant concerning the distance, since, as mentioned earlier, the systems characteristics are now fixed. This additional modelling gives $\theta = 85^{\circ}$ based on the fact that the jet precesses, or more precisely: "[...] the jets `wiggle' slightly about the best-fit position angle". The authors performed a detailed modelling, similar to what has been done by one of the authors on the galactic source SS433 which shows clearly jet precession \citep{Hjellming-Johnston-1981a,Hjellming-Johnston-1981b,Hjellming-Johnston-1988}.

The model of the wiggles assume, of course, that they are true, and are due to the kinematics only. In Fig.~3. of \citet{Hjellming-Rupen-1995} we see the variations of position of roughly 4 different ejectas (following the main solid lines only). It is obvious that few points perfectly follow the best-fit constant proper motion. But it is not obvious at all that they form a regular periodic pattern. Succeeding at modelling these wiggles does not mean we can interpret them as the signature of jet precession taking place in \gro, since the low number of points makes the fit poorly constrained. At that point, it should be clear enough that these supposed wiggles do not help to secure the distance. But there are additional problems attached to them.

In order to interpret wiggles as precession, one must ensure that the global motion follow lines of constant proper motion. But the 22 epochs of \emph{VLA} observations used for by this model did not resolve the source at a level of 100 mas, but only as a multi-core object elongating with time. The reason why a constant proper motion is a reasonable hypothesis is because it is consistent with what is seen in the \emph{VLBA} observations. But the authors emphasize the lack of a very-long-baseline interferometry calibrator, which implies that these \emph{VLBA} data are self-calibrated, "eliminating all absolute positional information, and leaving the alignment of the different images a free parameter." Consequently, the fact that the brightest point in each image is the stationary center of ejection is an hypothesis. 

Moreover, as mentioned in the paper, "these [VLA] and other [unspecified] data are consistent with constant intrinsic proper motion of 54 mas~d$^{-1}$" and later "the underlying proper motions appear constant". This value is in agreement with the result of equation~\ref{mumu4}. The hypothesis of constant proper motion is strengthened by the fact that the daily Southern Hemisphere VLBI Experiment (SHEVE) array observations of \citet{Tingay-etal-1995} are consistent with the major structures of the \emph{VLBA} images of Hjellming \& Rupen. As a matter  of fact, the authors note that the proper motion inferred from SHEVE data of 65$\pm$5 mas~d$^{-1}$ actually agrees with the 62 mas~d$^{-1}$ motion of the {\it outer edge} of the early NE ejecta. All these measurements indeed appear to roughly agree, but the uncertainties are likely large enough to encompass the wiggles. In summary, not only are the mere existence of these wiggles doubtful, but the simple ability to extract meaningful and additional constraints from it is also questionable. 

We conclude that the value of 3.2 kpc has been chosen between a firm upper limit of $3.5$~kpc and the external indication that its lower limit must be "about 3 kpc", which consequently fix the value of $\theta$. A questionable model is used to strengthen its value, and, as an obvious consequence, confirm the distance value. Literally, later in the text, it is said that the "[...] kinematic model for \gro\ {\it gives} a distance of 3.2$\pm$0.2 kpc". 

\subsection{Where does 3 kpc come from?}

As noted above, three references are given for a first estimation of the distance: \citet{Harmon-etal-1995}, \citet{Tingay-etal-1995} and \citet{McKay-Kesteven-1994}. Unfortunately, the first reference is literally citing the two others for the distance value, and must therefore be discarded. The paper by \citet{McKay-Kesteven-1994} is actually an IAU Circular which simply states that "HI observations of \gro\ made with the \emph{AT Compact Array} show solid absorption in the velocity range $+10$ to $-30$ \kms, with a further isolated weak feature at $-50$ \kms. {\it The balance of probabilities is that the distance is around 3.5 kpc, unless the $-50$ \kms\ feature is due to an atypical cloud.}" Although not being a robust measurement, the result obtained in this Circular needs to be verified.

Identifying the origin of the negative-velocity features in the absorption spectrum is the crucial point, since it provides an estimate of the lower limit of the distance, if correctly interpreted (i.e. if correctly identified and attributed to a component whose velocity can be estimated). To interpret the radio spectrum, one must consider a background source of unknown distance emitting continuum radiation (here \gro) that is being intercepted by foreground HI clouds. When one looks inside the solar galactic orbit, the line of sight goes through the multiple galactic spiral arms. Assuming all the clouds are moving with the mean galactic rotation scheme, the more distant from the Sun the cloud is, the more negative is its velocity, in the Local Standard of Rest, down to the tangential point where the distance/velocity relation flips back. 

\citet{Tingay-etal-1995} presented new radio \emph{VLBI} and \emph{ATCA} data of \gro. It is the only true work studying the lower distance limit of \gro. Firstly, we note that a rather large distance is expected by the authors in order to agree with: "a significant reddening due to absorption", as explained in \citet{dellaValle-1994}. \citet{dellaValle-1994} is a IAU Circular stating nothing more than: "[...] The [optical] spectrum exhibits prominent, broad Balmer lines [...] superimposed on a relatively red continuum. [...]" However, the spectrum has been taken during a flaring state. The spectral type of the secondary, and its possible veiling by the accretion disk, were, at that time, unknown. Leaving this aside, let us concentrate on the radio spectrum.

The HI spectrum of Tingay and collaborators, obtained with \emph{ATCA} (see their Fig.~2), shows a multi-component profile, with strong features at $\sim+5$\kms, between $-10$ and $-20$\kms, and isolated weak features at $-30$ and $-50$~\kms\ ($\sim$18\% and $\sim$2\% of the normalized continuum flux respectively). It is said that the latter feature is confirmed with multiple observations but no references are given. 

\begin{figure}
	\centering 
	\includegraphics[width=1.0\linewidth]{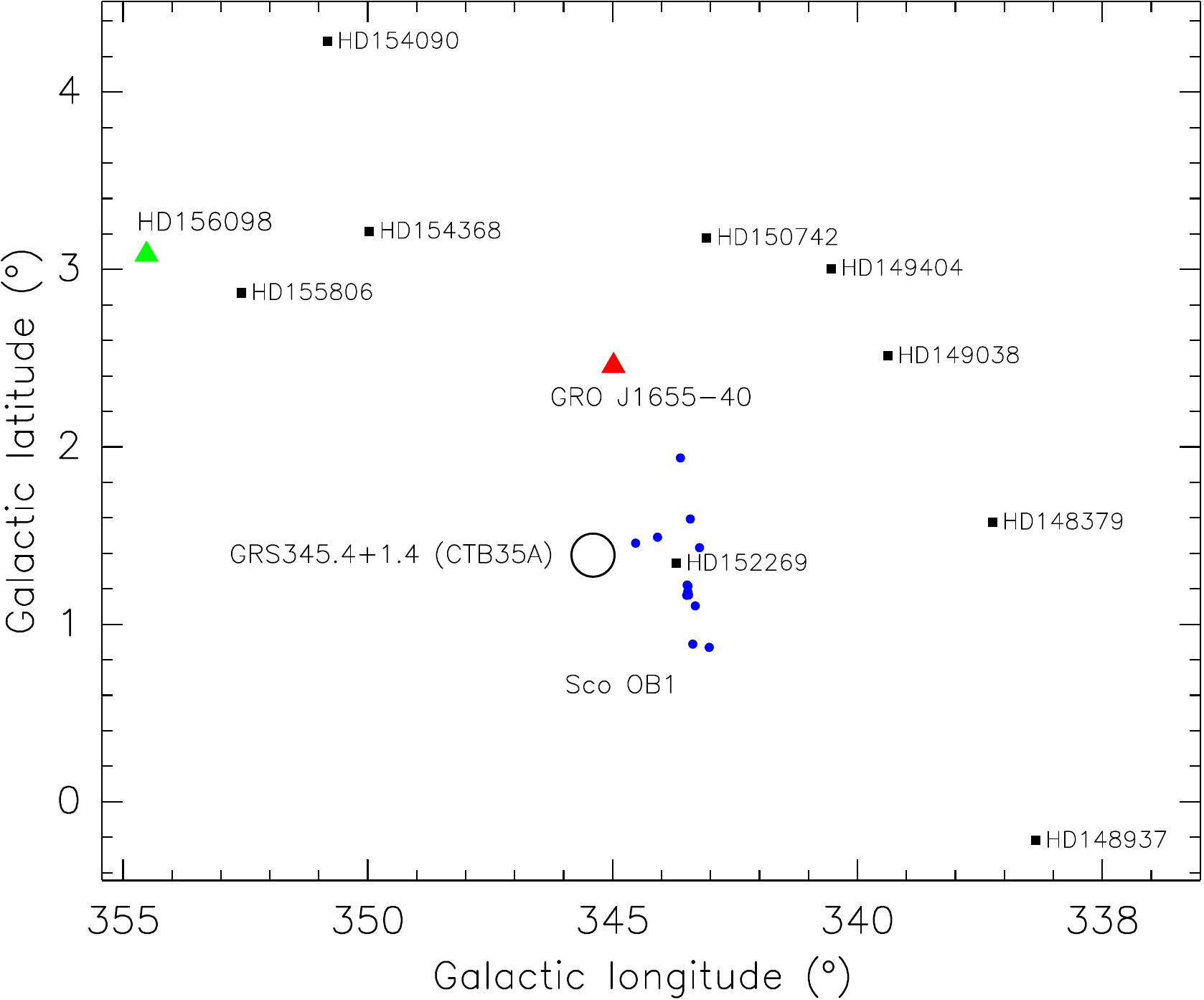} 
	\caption{Map of the region around \gro, where the region GRS345+1.4 studied by \citet{Caswell-etal-1975} is also shown as a big open circle. The 24 stars studied by \citet{Crawford-etal-1989} are indicated (stars belonging to Sco OB1 in blue circles, field stars in black squares with names). It shows that \gro\ is located at the core of the region studied by Crawford and collaborators. Note the presence of the star HD 152269 which lies close to the center of Sco~OB1 but is a foreground star. The green triangle to the left shows the position of the comparison F6IV star HD~156098 used by \citet{Foellmi-etal-2006b} in their analysis.} 
	\label{fig_mapScoOB1} 
\end{figure}

According to \citet{Tingay-etal-1995}, the feature at $-50$\kms\ would imply a lower limit of $\sim$4.2 kpc if it was participating to the mean galactic rotation\footnote{We assumed that the authors follow the same galactic model of \citet[][]{Caswell-etal-1975} which is not explicitly stated but whose results are explicitly used.}. This feature is actually discarded by Tingay and collaborators because it cannot be ruled out that such feature is driven by an expanding shell surrounding the Scorpius OB1 association located at 1.9 kpc from the Sun (see Fig.~\ref{fig_mapScoOB1}). This is however the feature used by \citet{McKay-Kesteven-1994} to derive an approximate distance of 3.5~kpc. In other words, \citet{McKay-Kesteven-1994} derive $D \sim 3.5$~kpc thanks to the feature at $-50$~\kms\ that is discarded by \citet{Tingay-etal-1995} because it would imply $D \gtrsim 4.2$~kpc.

According to \citet{Tingay-etal-1995}, the feature at $-30$\kms\ implies a lower limit of 3.0~kpc if it is associated with the mean galactic rotation. To strengthen their conclusion, they compare their spectrum with that of a nearby region GRS 345.4+1.4 (a.k.a. CTB~35~A) studied by \citet{Caswell-etal-1975} and located 2.4~kpc away. In fact, the lower limit of the distance of \gro\ is built on this hypothesis: the feature at $-30$\kms\ in the radio spectrum is moving with the mean galactic rotation scheme, which is confirmed by the comparison with GRS 345.4+1.4. 

But if there is such uncertainty on the interpretation of the feature at $-50$~\kms, why can't the feature at $-30$~\kms\ not also be associated with Sco~OB1? Or could the latter be indeed associated with Sco~OB1 while the most negative one is not? The spectrum of GRS~345.4+1.4 looks indeed similar to that of \gro, except that it has no absorption feature with velocities more negative than $-24$ \kms. But as shown on the map in Fig.~\ref{fig_mapScoOB1}, GRS345.4+1.4 is closer to Sco OB1 than to \gro. Why does the spectrum of GRS345.4+1.4 shows nothing more negative than $-24$~\kms\ although it is angularly closer but still behind the association compared to the position of \gro? 

We note moreover that \citet{Caswell-etal-1975} confirm the results of \citet{Radhakrishnan-etal-1972} that the HI absorption spectrum of the region GRS~345.4+1.4 is problematic since either the foreground absorbing cloud at $V = -24$ \kms or the HII region behind it (which is responsible for the continuum emission) has a "peculiar" motion of $\sim$10~\kms. \citet{Caswell-etal-1975} determine a (near kinematic) distance of 2.4 kpc by {\it assuming} a mean velocity of the multi-profile HI absorption of $\sim -20$\kms. An uncertainty of 10 \kms translates to an uncertainty of about 1.5 kpc in distance, if one uses the scale in \citet{Caswell-etal-1975}\footnote{We note that Caswell and coworkers use actually the galactic model of \citet{Schmidt-1965} with a galactic center distance of 10~kpc, while the modern value is about 8.0~kpc \citep{Groenewegen-etal-2008}. The extraction of the velocity-distance relationship directly on the figure of Caswell et al's paper reveals that the printed scale does not correspond exactly to the results written in the text itself (a velocity of $-24$~\kms correponds to 2.8~kpc instead of 2.4~kpc; possibly causing the confusion). However, the use of a smaller galactic center distance roughly compensate this error on the velocity-distance relationship.}. Furthermore, \citet{Shaver-etal-1982} have shown that there are a number of HI clouds in this region that have a peculiar velocity which cannot be accounted for by assuming that such cloud is moving with the mean galactic rotation, since the derived distance from the HI absorption appears larger than that derived from optical observations of HII regions along the same line of sight. We are therefore entitled to conclude that no meaningful comparison can be made between two absorption radio spectra in this region of the sky and the interpretation of the negative velocity features in the absorption radio spectrum is uncertain. 

We conclude that the lower limit of 3.0~kpc on the distance of \gro\ has not been established.

\subsection{Optical spectroscopy of stars in the direction of \gro.}

\citet{Crawford-etal-1989} have studied the interstellar sodium and calcium absorption lines towards the Scorpius OB1 association with high-resolution optical spectroscopy. The stars observed in this study are also shown in Fig.~\ref{fig_mapScoOB1}. They observed that all the spectra of Sco~OB1 members show structured features in the sodium doublet lines with velocities spanning a range as wide as 40 to 60 \kms, to the contrary of all other field stars, including the foreground star HD~152269 located right at the center of Sco OB1 on the sky. The absorption features at the bluer wavelengths inside the absorption lines are interpreted as truly blueshifted because the spectra are calibrated with atmospheric water lines directly on the spectra (i.e. the zero-point in velocity is known). 

\citet{Crawford-etal-1989} made the {\it global} following conclusions, despite variations between individual sources. Firstly, absorption lines comprised in the range $-20 \leq v_{\textrm{\scriptsize helio}} \leq 0$~\kms\ must arise from material between Sco~OB1 and the observer, since it cannot participate to the mean rotation curve. Second, the absence of blueshifted absorption features with $v_{\textrm{\scriptsize hel}} \leq -20$~\kms\ in the spectrum of HD~152269 implies that the absorbing material responsible for these blue features is comprised between the star (720 pc away), and the cluster. Given an average velocity of Sco~OB1 of about $-25$~\kms, the authors conclude that the absorptions seen at the bluest wavelengths of the sodium doublets are caused by expanding material associated with the cluster.

\begin{figure}
	\centering 
	\includegraphics[width=1.0\linewidth]{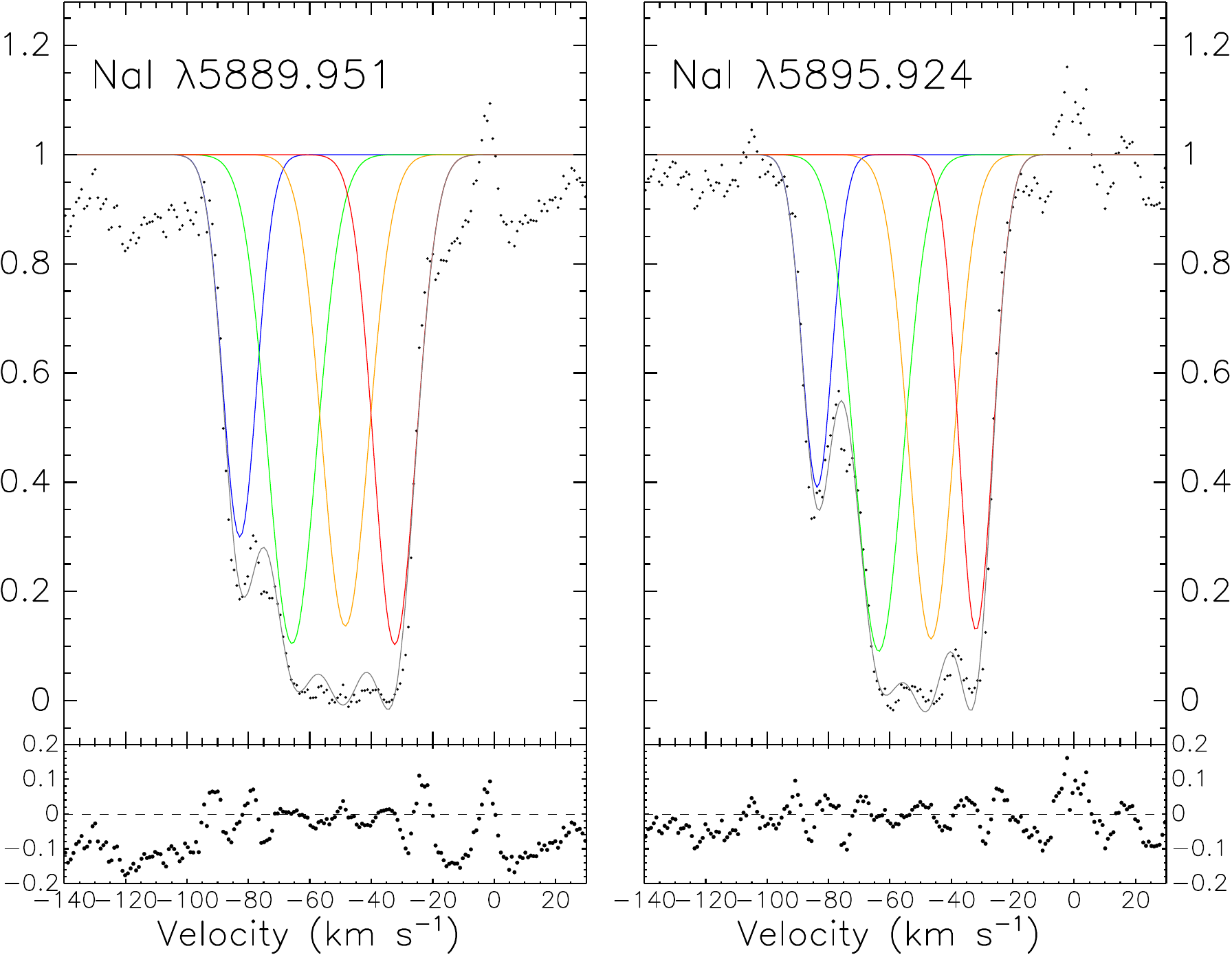} 
	\caption{The sodium doublet observed in the spectrum of \gro\ as function of the heliocentric velocity. Rest wavelength of the sodium were used: $\lambda\lambda$5889.951, 5895.924. The spectra were taken from \citet{Foellmi-etal-2006b}, and were combined together without preliminary radial-velocity shift. The zero-point of the velocity scale in the figure is simply that of the rest wavelength of the sodium lines, to which the calibration of the spectrum agree within 5~\kms\ (see text). Four gaussians were fitted to each independent line. The top panels show the spectra (points), with the individual gaussians in color, and the total gaussian in gray line. The bottom panels shows the difference between the spectrum and the fit.} 
	\label{fig_sodium} 
\end{figure}

In order to compare with \gro, we have taken the original spectra published in \citet{Foellmi-etal-2006b}, and averaged them with no radial-velocity shifts prior to the combination, in order to build a mean spectrum where the (static) interstellar lines are well averaged. The spectrum is shown in Fig.~\ref{fig_sodium}. Our UVES spectra are well aligned in a relative manner, since all sharp reddest wings of the Na lines appeared well aligned to each other. Moreover, we have identified about 10 (residual) sky lines inside the averaged spectrum (also identified in Foellmi et al. 2006). When compared to the skylines atlas of \citet{Osterbrock-etal-1996}, the velocity zero-point our the spectrum appears to be correct within 5~\kms. We have fitted the interstellar lines with four gaussians. We note the differences, in velocity, between the bluest and the reddest gaussians are 51 and 52 \kms\ respectively for the two lines. 

We can see in Fig.~\ref{fig_sodium} that there is no absorption between 0 and $\sim -30$\kms. Along with the conclusions of \citet{Crawford-etal-1989}, it would mean that there is a very small amount of absorbing material between \gro\ and the observer. This conclusion is hardly reconcilable with a large distance (and $a fortiori$ larger than that of Sco~OB1). Moreover, the sodium absorption is saturated between $-20$ and $-60$\kms\ and a weaker blue feature exist at $\sim -85$\kms. It seems more likely that these absorptions arise from an expanding material related to \gro\ itself. As matter of fact, \citet{Combi-etal-2007} have investigated the region of \gro\ with radio and infrared data tracing the HI, CO, and the gas morphology of the region. They have found that there is evidences of an HI hole of 1$^{\circ}$.5$\times$1$^{\circ}$.5 in diameter at a distance of $D = 1.2\pm 0.4$~kpc, and compressed CO material accumulated in part of the shell border, as well as infra-red emission with characteristics of shocked-heated dust. This study strengthen the suggestion that \gro\ could be located around 1.0~kpc, and possibly originated from NGC~6242. 

\section{The difficulties in estimating the extinction towards \gro}

After the value of 3.2 kpc had been published, a significant number of studies confirmed or strengthened this result, by using various distance methods. We discuss in the following sections the various problems encountered when estimating the extinction towards \gro. 

\subsection{The validity range of the sodium equivalent width-color excess relationship}
\label{sec_sodium}

A relationship exists between the equivalent width of the sodium doublet or the calcium lines in an optical spectrum, and its color excess. This relationship has been used, for instance, by \citet{Bianchini-etal-1997} for \gro, although using spectroscopy of low resolution preventing them to see the saturated and multi-profile nature of the lines \citep[see][for the details, and Fig.~\ref{fig_sodium} above]{Foellmi-etal-2006b}. They use the relationships given by \citet{Herbig-1975} for the calcium line at 6613\AA\ (see Fig.~4 of Herbig 1975) and \citet[]{dellaValle-Duerbeck-1993} \citep[who actually use the photoelectric photometry of ][]{Cohen-1975} for the sodium doublet at 5980\AA: 
\begin{equation}
	E(B-V) \sim 0.61 \times EW_{\textrm{\small NaI-D}} - 0.08 
	\label{equ_dellaValleDuerbeck}
\end{equation}

\citet{Bianchini-etal-1997} obtained a color excess ranging from 0.97 to 1.30 mag, from which they {\it adopt} a mean value of 1.13, and state that it is in agreement with the value found by other studies on this target \citep[for e.g.][but see below]{Bailyn-etal-1995a}. But such a range on the color excess translates, using a standard value of $R=3.1$, to an uncertainty of one magnitude on the absorption $A_V$ and therefore a factor of 1.6 in the distance.

Moreover, \citet{Munari-Zwitter-1997} have shown that the unambiguous range between equivalent widths of NaI lines and $E(B-V)$ is $0 \leq E(B-V) \leq 0.4$, i.e. much lower than the value measured by Bianchini and coworkers. If one used the (scattered) relation of \citet{dellaValle-Duerbeck-1993}, the limiting range of Munari \& Zwitter translates into equivalent widths between 0.13 and 0.79 \AA\, i.e. with an upper limit much lower than the value of 2.26~\AA\ from Bianchini et al. Moreover, assuming a single gaussian profile, it implies\footnote{FWHM$_{\textrm{\scriptsize Gauss}} = 2 \sqrt{2\ln 2} \cdot \sigma$ where $\sigma$ is the gaussian width.} a FWHM of the lines between 0.12 and 0.75 \AA, or a corresponding resolving power of about 50 000 for the narrowest Na lines, and 8000 for the broadest. It means that Bianchini and coworkers simply do not have the resolution necessary to derive an accurate value of the color excess to confirm the value of 3.2 kpc.

\subsection{Six extragalactic supernovae}

A distance "of $\sim$3.0~kpc" was proposed on the basis of optical data by \citet{Bailyn-etal-1995a}. The authors have also measured the equivalent widths of NaI-D lines in their spectrum (EW = 4.5\AA) that has a resolution of $\sim10$\AA\ \citep[which is even worse than that of][and moreover blended with HeI emission]{Bianchini-etal-1997}. They finally compute a color excess $E(B-V)$ = 1.15, using the relation between the equivalent width and the color excess given by \citet{Barbon-etal-1990}. The latter paper studies the type-Ia supernova SN1989B in NGC 3627. \citet{Barbon-etal-1990} determined an empirical and roughly linear relation between the equivalent width of the NaI-D lines and the color excess $E(B-V)$ with the spectra of (only) six extragalactic supernovae. Their relation reads (see the end of their section \S 3.2): 
\begin{equation}
	E(B-V) \sim 0.25 \times EW_{\textrm{\small NaI-D}} 
	\label{6sn} 
\end{equation}
defined for $E(B-V)$ between 0.1 and 1.0, after removing the galactic contribution. Leaving aside the intrinsic difficulty of determining the galactic contribution, the source of the data used to determine this relationship is not given in \citet{Barbon-etal-1990}, and we can question the reliability of a relation calibrated with 6 points only. Moreover, using the relation given by \citet{dellaValle-Duerbeck-1993}, equation~\ref{equ_dellaValleDuerbeck} would give $E(B-V)$ = 2.66, instead of 1.15. It implies an absorption value of $A_V^{\textrm{\scriptsize opt}} \sim 8.24$ instead of 3.56, and therefore a factor of 8.7 in the relative value of the distance (equation~\ref{equ_magdist}). Even ignoring the line saturation in the case of \gro, the two relationships are still hardly reconcilable.

Interestingly, \citet{Bailyn-etal-1995a} claimed that their result "is consistent with the EW of other interstellar lines in the optical domain". Finally, they use the classical relationships of \citet[][discussed below]{Allen-1973} and \citet[][who actually did not study the sodium doublet]{Herbig-1975}, to conclude that the distance of the source is compatible with $D \sim 3$ kpc, "in agreement with the radio observations" of \citet{Tingay-etal-1995}.

\subsection{Where is the \emph{HST/STIS} spectrum?}
\label{sec_hst_stis}

\citet{Orosz-Bailyn-1997} presented an extended spectroscopic and photometric dataset. In particular, they have found very clear ellipsoidal variations in their \emph{BVRI} lightcurves, obtained in February and March 1996, when the system was not yet completely in quiescence. Although they mention various consistency checks throughout the paper, they do not measure the distance, but rather rely on that of \citet{Hjellming-Rupen-1995}, said, along with $E(B-V)$, to be "tightly constrained". As for the color excess, they assumed a value of $E(B-V) = 1.3 \pm 0.1$, actually obtained by \citet{Horne-etal-1996} who used high-quality UV spectra obtained with the \emph{Hubble Space Telescope}. However, \citet{Horne-etal-1996} is an IAU circular where the spectrum is not visible, and where it is simply stated that "deep 220-nm absorption in the HST spectrum {\it suggests} $E(B-V)$ = 1.3 mag." 

Similarly, \citet{vanderHooft-etal-1998} presented new $VRi$ photometric data acquired during 28 consecutive nights in March 1996 with the Dutch 0.91m telescope in La Silla (Chile), when the source was said to be close to its quiescence brightness (but see below). From the modelling of the lightcurve they obtain a inclination angle of about 67$^\circ$ and a black hole mass between 6.3 and 7.6 $M_{\odot}$ \citep[consistent with][]{Orosz-Bailyn-1997}. However, they have assumed again the radio distance of 3.2~kpc by \citet{Hjellming-Rupen-1995}, and a color excess of $E(B-V) = 1.3$ mag, taken from \citet{Horne-etal-1996}. Interestingly, they mention that the distance is well constrained since it has been found consistent with the results of \citet{McKay-Kesteven-1994}, \citet{Tingay-etal-1995}, \citet{Bailyn-etal-1995a} and \citet{Greiner-etal-1995}. We have shown above how weak were these first three references concerning the distance of \gro\ (the fourth and last reference is discussed in Sec.~\ref{sec_X-rays}). 

\subsection{A new and lower value of the color excess of \gro}

One can directly estimate the color excess of \gro. As mentioned in \citet{Foellmi-etal-2006b}, the F6IV comparison star HD~156098 has a known Hipparcos distance of 50$\pm$0.2 parsec. It is close enough to assume that it has a negligible interstellar absorption \citep[see for instance][]{Welsh-etal-1990}. Therefore we assume that its observed color index is equal to its intrinsic color index: $(B-V)_{F*} \equiv (B-V)_{F*,0}$ = 0.46 [using SIMBAD; see also \citet{Fitzgerald-1970} who give $(B-V)=0.46$ for an F6IV]. Using the mean visual magnitude $V = 17.12$ from \citet{Orosz-Bailyn-1997} and adopting $B \sim 18.6$\footnote{There is an error in \citet{Foellmi-etal-2006b} who use $B\sim16.65$ instead of $B\sim18.6$.} from their lightcurve, the color index of \gro\ reads: $(B-V)_{\textrm{\tiny GRO}} = 1.48$~mag. Since \citet{Foellmi-etal-2006b} have shown that HD~156098 represents a fairly good twin to the spectrum of \gro\footnote{We emphasize here that the {\it spectra} look the same, i.e. the global average spectroscopic parameters are comparable. We obviously do not mean that the {\it stars} are identical. This distinction is important in the next sections.}, the color excess of the microquasar simply follows: $E(B-V) = (B-V)_{\textrm{\tiny GRO}} - (B-V)_{F*,0} = 1.02$~mag, which is smaller than previous values. This conclusion is also reached by \citet{Beer-Podsiadlowski-2002} who mention that the value $E(B-V)=1.3$ is not consistent with a F6IV star but rather an A8 or earlier. The value of $(B-V)_{F*,0}$ is a bit outside the validity range of equation~\ref{equ_R}. If we nevertheless use this equation, we derive $R = 3.42$ and therefore $A_V = 3.49$. The implications of these new values to the luminosity and mass ratio of the \gro\ are discussed in Sec.~\ref{sec_dynamics}.

\section{The optical absorption derived from X-ray or optical data}
\label{sec_X-rays}

We have seen above some difficulties at determining the optical absorption towards a source. This {\it optical} absorption can in practice also be obtained from {\it X-ray} data. We present here the issues related to that approach and its application to \gro. In particular, we show that no systematic overestimation from X-ray data when compared to optical data can be claimed.

\subsection{Is the optical absorption of \gro\ from \emph{ROSAT} flawed?}

\citet{Greiner-etal-1995} presented new \emph{ROSAT} X-ray data of \gro, from which they infer a distance of 3~kpc (no uncertainties are provided). Their method consists of fitting the halo of the observed radial profile of the source. This halo is produced by the scattering of the X-rays by the interstellar dust. To fit the radial profile of \gro\ observed with \emph{ROSAT}, they assume an uniform dust distribution between the observer and the source in their model. They obtain, with not many details, a value of the effective optical depth at 1~keV of $\tau_\mathrm{eff} \sim 0.33$.

Furthermore, they use the results of \citet{Predehl-Schmitt-1995} who have studied in details X-ray halos in \emph{ROSAT} sources, and have shown that a good correlation exists between the simultaneous measured dust and hydrogen column densities: "indicating that gas and dust must be to a large extent cospatial". From the fractional halo intensity it is thus possible to derive the dust column density. The relation used by \citet{Predehl-Schmitt-1995} reads (see the end of their Sec. 3.4): 
\begin{equation}
	\label{NH_AV} \tau_{\mathrm{eff}} / A_V= 0.056 \pm 0.01 
\end{equation}
where $A_V$ is the visual absorption (expressed in magnitude).

Using this correlation, and assuming that the sight-line for \gro\ has the same gas-to-dust ratio as the sight-lines for which the relations between $A_V$ and $N_H$ have been established, \citet{Greiner-etal-1995} obtain for \gro\ an absorption of $A_V = 5.6$ mag (and a hydrogen column density of $N_H = 7.0 \times 10^{21}$ cm$^{-2}$, implying a color excess of $E(B-V)\sim$1.7, assuming $R = 3.1$). This is significantly (by $\sim 1.5 - 2$ magnitudes) larger than any other estimation from the studies in the optical ($A_V \sim 3.8-4.1$ mag) and our own estimate above.

To finally compute the distance, \citet{Greiner-etal-1995} use the mean extinction law given by \citet[Sec. 125, p.263]{Allen-1973}: $A_V = 1.9$ mag~kpc$^{-1}$, and say that it is in agreement with other determinations of the distance. It is however hard to consider the relation of Allen more than a basic approximation, since it is composed of two parts: $A_V = 1.6$ mag~kpc$^{-1}$ from "interstellar absorbing clouds" and $A_V = 0.3$ mag~kpc$^{-1}$ from "grains between the clouds". The value of 1.6 mag~kpc$^{-1}$ is actually obtained by simply assuming that there are 5 clouds per kpc in the galactic plane, and that the mean visual absorption is about 0.3 mag per cloud \citep[Sec. 124, pp.262-263]{Allen-1973}. The value of 0.3 mag~kpc$^{-1}$ is quoted from \citet{Gottlieb-Upson-1969} who in fact clearly show, by dividing the sky into more than 200 zones, that the extinction in a particular direction is by far more complicated than such a simple mean relation. 

In summary, the value of the optical absorption determined by \citet{Greiner-etal-1995} is intriguing and significantly larger than what we obtained above: $A_V = 3.49$. On the other hand, their "confirmation" of the distance cannot be trusted, even if, mentioning the 3-5 kpc range of \citet{Tingay-etal-1995}, they claim in their conclusions that: "The scattering of X-rays by the interstellar dust allows to derive a distance of \gro\ of 3 kpc." 

\subsection{Is there a systematic effect?} \label{Ax_Av}

The significantly larger value of \avx\ compared to \avo\ is important since it can provide strong insights into the role of dust, and the possible superabundance of dust over hydrogen close to the source itself. It is not only an important point for the distance determination, but also for our understanding of the closeby environments of such objects and therefore deserves to be verified.

\citet{Jonker-Nelemans-2004} claim that optical absorption obtained from X-rays are systematically larger than those obtained from optical data, as shown in their Table 3, in which the values of \avo\ and \avx\ are listed for 14 sources, among which \gro\ and \aaa. Taken at face value, four targets in the list of 14 sources don't have a value in one of the two bands (an "X" is marked in its place): GS~1009-45, XTE~J1118+480, H~1705-250, SAX~J1819.3-2525. Moreover, five others have, strictly speaking, consistent values either because the values agree within the (large) uncertainties or because only upper/lower limits were determined (GRO~J0422+32, \aaa, \gro, GX~339--4 and GS~2~Section 023+338), leaving 5 systems only for a true comparison: GS~1124--684, 4U~1543--47, XTE~J1550--564, XTE~J1859+226 and GS~2000+25. 

\subsubsection{GS~1124--684}

For GS~1124--684 (=GU Mus, Nova Mus 1991), \citet{Cheng-etal-1992} determine the color excess to be $E(B-V) \sim 0.29$, which corresponds to \avo$=0.9\pm0.1$, by fitting a model of a $HST/FOS$ spectrum between 1600 and 4900\AA.

The value of \avx$=1.28\pm0.06$ is computed by \citet{Jonker-Nelemans-2004} taking $N_{H} = 2.28 \times 10^{21}$ cm$^{-2}$ of \citet{Greiner-etal-1994a} and using the relation $N_{H}/A_V = 1.79 \times 10^{21}$ cm$^{-2}$ by \citet{Predehl-Schmitt-1995}\footnote{It is actually misspelled "Schmi{\it dt}" in the caption of Table~3 of Jonker \& Nelemans.}, corresponding to a color excess of $E(B-V) \sim 0.4$. But \citet{Greiner-etal-1994a} write that their estimation of the color excess is confirmed by the works of, first, \citet{dellaValle-etal-1991}, and second, \citet{Cheng-etal-1992} which is the reference for \avo. Moreover, \citet{dellaValle-etal-1991} use the problematic 6-extragalactic supernovae relation of \citet{Barbon-etal-1990} discussed above and adopt a mean value for the color excess derived from a {\it range} of values between 0.2 and 0.35, i.e. in agreement with that of \cite{Cheng-etal-1992}. 

As mentioned by \citet{Greiner-etal-1994a} "the reddening of the X-ray transient in Muscae was first derived to $E(B-V) \approx 0.2-0.3$ from IUE and optical measurements", and give three references, among which two IAU Circulars by \citet[][but see also Appendix~\ref{graph}]{West-etal-1991}, and \citet{Gonzalez-Riestra-etal-1991}. Unfortunately, none of the two circulars report a value of the color excess as such. The third reference given is a conference proceeding of Shrader C. and Gonzalez-Riestra R. 1991 (in "Workshop on Nova Muscae 1991", Lyngby 1991, DRI Prep. 2-91, p.85) which is unfortunately not referenced in the NASA ADS system. 

We can safely discard the source GS~1124--684 in Table~3 of \citet{Jonker-Nelemans-2004}.

\subsubsection{4U~1543--47}

For 4U~1543--47 (=IL Lup*), the optical value $A_V=1.55\pm0.15$ is quoted from \citet{Orosz-etal-1998} who also mention that their value is smaller than that derived from X-rays, quoting the values of \citet{Greiner-etal-1994b} and \citet{vanderWoerd-etal-1989} which is the reference for \avx used by \citet{Jonker-Nelemans-2004}.

Strictly speaking, \citet[][see their Sec. 4]{Orosz-etal-1998} have the following range $0.45 < E(B-V) < 0.55$, while the values of the color excess obtained from X-ray data ranges from "0.56 to 0.77 assuming $A_V=N_H/1.79 \times 10^{21}$ \citep{Predehl-Schmitt-1995}." That is, the two ranges almost agree. \citet{Jonker-Nelemans-2004} ignore the value of \citet{Greiner-etal-1994b} and choose the upper limit to derive \avo$ = 2.4\pm0.1$. Given the difficulties of obtaining reliable estimates in both energy regimes, this disagreement between \avo\ and \avx\ is not really conclusive.

\subsubsection{GS~2000+25}

For GS~2000+25 (=QZ Vul, Nova Vul 1988), the value from X-ray observations quoted by \citet{Jonker-Nelemans-2004} is $A_V = 6.4\pm1.0$, citing \citet{Tsunemi-etal-1989}. However, in this latter paper, the value quoted is $A_V = 4.41$ (or $\log_{10} N_H = 22.06\pm0.006$, see their Sec. III, a, iii), said to be in agreement with the optical estimation by \citet[][\avo\ = 3.5]{Chevalier-Ilovaisky-1990} which is the reference given by Jonker \& Nelemans for the optical value, who also note the large uncertainty of the latter. This source must therefore also be discarded from Table~3.

\subsubsection{XTE~J1550--564 and XTE~J1859+226}

As for the two remaining sources, namely XTE~J1550--564 (=V381 Nor) and XTE~J1859+226 (=V406 Vul), there seems to be a difference between \avx\ and \avo, but not necessarily because of a systematic effect in one of the two wavelength regime.

For the X-ray binary XTE~J1550--564, \citet{Jonker-Nelemans-2004} quote \avo = 2.5 (no uncertainties) computed from \citet{Sanchez-Fernandez-etal-1999} using their value of the equivalent width of the Diffuse Interstellar Band (EW=1.9\AA, for the 4430\AA\ DIB) and the relation of \citet{Herbig-1975}. \citet{Sanchez-Fernandez-etal-1999} actually report \avo=2.2 from $\langle E(B-V)\rangle = 0.7\pm0.2$ using a combination of the NaI lines and the DIB and using again the 6-supernovae relationship of \citet{Barbon-etal-1990} discussed above. Given the uncertainties on $E(B-V)$, the two values actually agree. But if one uses the relationship of \citet{dellaValle-Duerbeck-1993}, $E(B-V) = 1.6$, and \avo $\gtrsim$ 5. In any case, their equivalent width of the NaI lines of 2.4\AA\ is much larger than the significant range computed above from the study of \citet{Munari-Zwitter-1997}: $EW_{\mathrm{NaI}} \approx 0.1 - 0.8$\AA\ (or $0. \leq E(B-V) \leq 0.4$). Moreover, they use a low-resolution (2\AA) spectrum, and it is not clear if the lines are saturated or not, which is the reason why \citet{Jonker-Nelemans-2004} use their value of the DIB only.

As for XTE~J1859+226, \citet{Jonker-Nelemans-2004} mention that \avx=4.47 has no error bars. This value is quoted from the IAU Circular no. 7291 by \citet{dalFiume-etal-1999} who found N$_H$ to be "about $8 \times 10^{21}$ cm$^{-2}$". \avo=$1.80\pm0.37$ is given by \citet{Hynes-etal-2002} who found $E(B-V) = 0.58\pm0.12$ from a spectral fit of the UV feature at 2200\AA\ in their \emph{HST/STIS} spectrum. Assuming \avx\ being correct, the 2.6 mag difference is certainly due to local absorption in the source itself. 

\subsection{Is there a discrepancy between \avx\ and \avo?}

To summarize, the claimed {\it systematic} overestimation of the optical absorption obtained from X-ray observations compared to the value obtained from optical studies is not established, since only two sources (XTE~J1550--564 and XTE~J1859+226) among the 14 used by \citet{Jonker-Nelemans-2004} potentially show any difference. To claim a systematic effect in one given direction also implies the knowledge of where the correct value is. 

The overestimation of \avx\ over \avo\ mentioned by \citet{Jonker-Nelemans-2004} has  apparently already been observed by \citet{Vrtilek-etal-1991}. But Vrtilek and coworkers obtained {\it all} their absorption values from optical data from an older source: \citet{Bradt-etal-1983}, who do not always measure the absorption themselves but cite even older sources. Moreover, the supposed systematic overestimation of \citet{Vrtilek-etal-1991} is pictured in their Fig.~6a where the non-zero slope showing the effect is caused by 2 deviating points only. According to their own caption, one point is a high-mass X-ray binary (4U 1516-56, a.k.a. Cir~X-1) that we know today contains local absorption \citep{Johnston-etal-2001} and where near infrared (NIR) spectra in the K band revealed itself to be too obscured to allow a spectral classification \citep{Clark-etal-2003} \citep[see also][for HST data on this source]{Mignani-etal-2002}. The second deviating point of the figure is a low-mass X-ray binary (4U 1728-337, a.k.a. GX 354+00) in a globular cluster. The \avo\ of this source is determined by \citet{Grindlay-Hertz-1981} not with the mentioned optical data but rather with $JHK$ photometry. Moreover, the values of \avo\ and \avx\ for this object agree within 0.2 mag. 

We note that two sources (namely XTE~J1859+226 and \gro) seem to have \avx\ significantly larger than \avo. Given that most of the X-ray observations are done during outbursts, and it is tempting to think that a local absorber makes the estimation of \avx\ flawed. We note however that the work of \citet{Predehl-Schmitt-1995}, used by many for their relation between $A_V$ and $N_H$, show precisely that there is a surprisingly good correlation between the simultaneously measured dust and gas density. 

It seems clear that this point should deserve more studies, especially with new methods of determining \avx. For instance, \citet{Xiang-etal-2007} use time delays between photons that are being scattered by dust, and being recorded in the X-ray halo. Interestingly, these delays are also function of the distance to the source.

\subsection{The quiescence state might be variable}

In \citet{Jonker-Nelemans-2004}, the quoted value of the absorption of \gro\ is \avx$=4.8\pm2.8$, which is still in agreement both with the peculiar value of \citet{Greiner-etal-1995}, and the value of \avo\ from \citet[][\avo=$3.7\pm0.3$]{Hynes-etal-1998} and our own determination of \avo=3.49 above. Nevertheless, one additional source of uncertainty is whether the object was observed in quiescence. \citet{Jonker-Nelemans-2004} refer to \citet{Kong-etal-2002} concerning \gro. These authors found a value of the X-ray luminosity measured with \emph{Chandra} lower than that of \citet{Greiner-etal-1995} with \emph{ROSAT}, and even lower than another measurement made between two large outbursts separated by one year, obtained with \emph{ASCA} by \citet{Ueda-etal-1998} and \citet{Asai-etal-1998}. Kong and coworkers note that their 
\emph{Chandra} observation occurred about 4 years after the last outburst, and may therefore measure the quiescent X-ray level more accurately (although this is not necessarily true). In addition, \citet{Garcia-etal-2001} emphasize that \citet{Wagner-etal-1994} observed variations of the {\it quiescent} X-ray flux in V404 Cyg by a factor of 10 with \emph{ROSAT}. 

Finally, the variability in the optical and near-infrared has been recently demonstrated by \citet{Cantrell-etal-2008} for when the object is considered in X-ray quiescence. They have studied optical and infrared photometry of \aaa\ spanning the years 1999-2007. They clearly demonstrate the existence of three different states (passive, loop, active) in which only the passive one truly represent quiescence. Therefore, great care must be used when modelling optical lightcurves, since it will have a direct influence on the observed quiescent visual magnitude, and the determination of the Roche-lobe effective radius. These issues are discussed in the next section.

\section{The characterisation of the orbit and the secondary star of \gro}
\label{sec_dynamics}

We have shown the various difficulties at determining a robust value of the color excess and/or the absorption value. We now turn to the other problematic term in equation~\ref{equ_magdist}: the determination of the absolute magnitude of the secondary star, emphasizing that none of the studies mentioned in this section were aware of the requirement to isolate true {\it optical} quiescent lightcurves, as mentioned above.

\subsection{Comparison of height different studies of the orbital and secondary star parameters of \gro.}

To obtain the absolute magnitude of the secondary star, we need to combine information on the temperature of the star and information on its radius (see equation~\ref{equ_lum}). There are various problems arising with this approach, which are all exemplified in Table~\ref{orbital-parameters} where we extracted the quantitative results on the orbital parameters of \gro\ and its secondary star obtained by \citet{vanderHooft-etal-1997}, \citet{Orosz-Bailyn-1997}, \citet{vanderHooft-etal-1998}, \citet{Phillips-etal-1999}, \citet{Shahbaz-etal-1999}, \citet{Greene-etal-2001}, \citet{Beer-Podsiadlowski-2002} and \citet{Shahbaz-2003}. All these studies use either optical spectra or CCD photometric lightcurves in the optical domain, or both. Some reanalyse data published previously in order to correct a specific assumption in the given analysis. 

In Table~\ref{orbital-parameters} we have extracted the relevant orbital and secondary star parameters, being very careful to choose the best quantitative results according to the authors. In the lower part of the table, we have, in addition, computed the projected rotational $v \sin i$ (in \kms) using the relationship of \citet{Horne-etal-1986} when both the mass ratio $q \equiv m_2/m_1$ (where $m_1$ is the mass of the black hole; in Table~\ref{orbital-parameters} we use the inverse definition: $Q \equiv m_1/m_2$) and the secondary radial-velocity semi-amplitude $K_2$ were available\footnote{We note that \citet{Shahbaz-etal-1999} and \citet{Shahbaz-2003} also use this relationship but there is a typo in their formula: the 1/3 exponent is misplaced.}:
\begin{equation}
	v \sin i = K_2 (1+q) \frac{0.49 q^{2/3}}{0.6 q^{2/3} + \ln(1+q^{1/3})}
	\label{equ_vsini}
\end{equation}
We emphasize that this relationship is valid when the hypothesis of synchronous rotation is verified. When the inclination angle was available, we computed the rotational velocity $v_{rot}$. The Roche-lobe effective radius (in $R_{\odot}$) is calculated using the formula of \citet{Paczynski-1971} (see also \citet{Jonker-Nelemans-2004} for a discussion) from the period $P$ and the secondary mass $m_2$:
\begin{equation}
	R_2 \equiv R_{\textrm{{\scriptsize Roche Lobe}}} = 0.234 \; P_{\textrm{{\scriptsize orb}}}^{2/3} m_{2}^{1/3}
	\label{equ_RRoche}
\end{equation}
with $P$ the orbital period expressed in {\it hours} and $m_2$ in solar masses. Given the Roche radius and the orbital period, we computed the equatorial velocity of a star with the same radius, which allow a cross-check of the synchronous rotation hypothesis. 

When a temperature was available, we also computed the so-called Roche luminosity, using equation~\ref{equ_lum}, the Roche radius and assuming a solar temperature $T_{\odot} = 5800 K$. We then calculated the absolute magnitude $M_V$ from this Roche luminosity, using the absorption $A_V=3.49$ determined above for \gro, and the apparent magnitude $m_V$ from the considered data. Finally, we also computed the corresponding distance.

\begin{sidewaystable*}
	\centering \scriptsize
	\begin{tabular}{l r@{\,$\pm$\,}l r@{\,$\pm$\,}l r@{\,$\pm$\,}l c r@{\,$\pm$\,}l r@{\,$\pm$\,}l r@{\,$\pm$\,}l r@{\,$\pm$\,}l r@{\,$\pm$\,}l} \hline 
	
		Parameter 									&\multicolumn{2}{c}{vdH97$^{\mathbf{a}}$}
															&\multicolumn{2}{c}{OB97} 
															&\multicolumn{2}{c}{vdH98}
															& PSP99
															&\multicolumn{2}{c}{S99}
															&\multicolumn{2}{c}{GBO01}
															&\multicolumn{2}{c}{BP02}
															&\multicolumn{2}{c}{S03} \\ \hline 

		Data 											&\multicolumn{2}{c}{$BVRi$ lightcurves}
															&\multicolumn{2}{c}{Optical spec. \& $BVRI$ light.}
															&\multicolumn{2}{c}{$VRi$ lightcurves} 
															& OB97
															&\multicolumn{2}{c}{Optical spectra}
															&\multicolumn{2}{c}{$BVIJK$ lightcurves}
															&\multicolumn{2}{c}{OB97}
															&\multicolumn{2}{c}{S99}\\

		Observation dates					&\multicolumn{2}{c}{May--July 1995}
															&\multicolumn{2}{c}{May 1995, Feb. 1996}
															&\multicolumn{2}{c}{March--April 1996}
															& May 1995 only
															&\multicolumn{2}{c}{May--June 1998}
															&\multicolumn{2}{c}{July--October 1999}
															&\multicolumn{2}{c}{--}
															&\multicolumn{2}{c}{--}\\

		Orbital period ($d$) 			& 2.620040	& 0.00098	
															& 2.62157 	& 0.00015
															& 2.62168 	& 0.00014
															&vdH98
															&\multicolumn{2}{c}{vdH98}
															& 2.62191 & 0.0002
															&\multicolumn{2}{c}{OB97}
															&\multicolumn{2}{c}{vdH98 \& OB97$^{\mathbf{b}}$}\\

		RV amplitude K$_2$ (km\,s$^{-1}$)
															&\multicolumn{2}{c}{--}
															& 228.2	& 2.2
															&\multicolumn{2}{c}{--}
															& 196 $^{\mathbf{c}}$
															& 215.5	 &	2.4	
															&\multicolumn{2}{c}{S99}
															&\multicolumn{2}{c}{OB97}
															&\multicolumn{2}{c}{S99}\\

		Mass ratio $Q$ 						&\multicolumn{2}{c}{3.8--5.5}
															& 2.99 & 0.08
															&\multicolumn{2}{c}{2.43--3.99$^{\mathbf{d}}$}
															& OB97
															&\multicolumn{2}{c}{2.293--2.967}
															& 2.6 & 0.3
															& 3.9 & 0.6 
															& 2.386 & 0.028\\

		Inclination angle $i$ ($^{\circ}$)
															&\multicolumn{2}{c}{65--76}
															& 69.50 & 0.08
															&\multicolumn{2}{c}{63.7--70.7}	
															& vdH98
															&\multicolumn{2}{c}{vdH98}	
															& 70.2  & 1.9
															& 68.65 & 1.6
															&\multicolumn{2}{c}{OB97$^{\mathbf{e}}$}\\

		Black-hole mass $m_1$ ($M_{\odot}$)	
															&\multicolumn{2}{c}{4.94--6.79}
															& 7.02 & 0.22
															&\multicolumn{2}{c}{6.29--7.60}	
															& 4.1--6.7
															&\multicolumn{2}{c}{5.5--7.9}
															& 6.3  & 0.5
															& 5.40 & 0.30
															& 6.59 & 0.45 \\

		Secondary mass $m_2$ ($M_{\odot}$)
															&\multicolumn{2}{c}{0.96--1.78}
															& 2.4 & 0.5
															&\multicolumn{2}{c}{1.6--3.10}
															&1.4--2.2
															&\multicolumn{2}{c}{1.7--3.3}
															& 2.4  & 0.4
															& 1.45 & 0.35
															& 2.76 & 0.33 \\

		Vis. luminosity $L_2$ ($L_{\odot}$)
															&\multicolumn{2}{c}{--}
															& 46.6 & 13.6
															&\multicolumn{2}{c}{41 (fixed)$^{\mathbf{f}}$}
															& OB97
															&\multicolumn{2}{c}{--}
															&\multicolumn{2}{c}{--}
															& 21.0 & 6.0
															&\multicolumn{2}{c}{--}\\

		Temperature $T_{eff}$ ($K$)
															&\multicolumn{2}{c}{--}
															&\multicolumn{2}{c}{6500 (fixed)}	
															&\multicolumn{2}{c}{6330--6620}
															&6500 (fixed)$^{\mathbf{g}}$
															&\multicolumn{2}{c}{--}
															&\multicolumn{2}{c}{6336$^{\mathbf{h}}$}
															& 6150 & 350
															&\multicolumn{2}{c}{6600}\\

		Color excess $E(B-V)$ 			&\multicolumn{2}{c}{--}		
															& 1.3 				& 0.1$^{\mathbf{i}}$
															&\multicolumn{2}{c}{1.3$^{\mathbf{i}}$}
															&--
															&\multicolumn{2}{c}{--}
															&\multicolumn{2}{c}{--}
															& 1.0 & 0.1
															&\multicolumn{2}{c}{--}\\ \hline 

		Proj. rot. vel. $v \sin i$ (km\,s$^{-1}$)
															&\multicolumn{2}{c}{--}
															& 88.1 & 2.1
															&\multicolumn{2}{c}{--}
															& 	75$\pm$5
															&\multicolumn{2}{c}{83.5--95.9}
															& 89 & 7$^{\mathbf{j}}$				
															& 77 & 1
															& 93.8 & 0.6 \\

		Rotational vel. $v_{rot}$ (km\,s$^{-1}$)
															&\multicolumn{2}{c}{--}
															& 94 & 6.6
															&\multicolumn{2}{c}{--}
															& 80.01--84.24
															&\multicolumn{2}{c}{88.5--107.0}
															& 95 & 7
															& 83 & 1
															& 100 & 2 \\
															
		Roche Radius ($R_{\odot}$)	&\multicolumn{2}{c}{3.65--4.48} 							
															& 4.95 & 0.15
															&\multicolumn{2}{c}{4.3--5.4}
															& 4.1--4.8
															&\multicolumn{2}{c}{--}
															& 4.9 & 0.3
															& 4.2 & 0.2
															& 5.2 & 0.3\\

		Equ. Roche vel. (km\,s$^{-1}$)
															&\multicolumn{2}{c}{70.5--86.5}
															& 95.7 & 5.3
															&\multicolumn{2}{c}{83.6--104.2}
															& 80.0--93.0
															&\multicolumn{2}{c}{--}
															& 95 & 8
															& 80.9 & 6.6
															& 100.2 & 4\\

		Roche vis. luminosity ($L_{\odot}$)	
															&\multicolumn{2}{c}{21.0--31.6}
															& 38.6 & 2.3
															&\multicolumn{2}{c}{26.2--49.5}
															& 26.5--36.3
															&\multicolumn{2}{c}{--}
															& 34.9 & 1.2
															& 22.3 & 5.4
															& 45.3 & 5.2\\
																														
		Absolute mag. $M_V$ (mag)	&\multicolumn{2}{c}{1.5--1.6$^{\mathbf{k}}$}
															& 0.8 & 0.1
															&\multicolumn{2}{c}{0.6--1.3}
															& 0.9--1.3
															&\multicolumn{2}{c}{--}
															& 1.0 & 0.1
															& 1.5 & 0.3
															& 0.7 & 0.3\\
															
		Distance $D$ (kpc)		 			&\multicolumn{2}{c}{--}
															& 3.8 & 0.2
															&\multicolumn{2}{c}{1.9--2.7}
															& 3.0--3.6$^{\mathbf{l}}$
															&\multicolumn{2}{c}{--}
															& 3.4 & 0.2$^{\mathbf{l}}$
															& 2.86 & 0.45
															& 4.0 & 0.5 \\ \hline 
															
	\end{tabular}
	\caption{Comparison table between various parameters of GRO J1655-40 as measured by \citet[vdH97]{vanderHooft-etal-1997}, \citet[OB97]{Orosz-Bailyn-1997}, \citet[vdH98]{vanderHooft-etal-1998}, \citet[PSP99]{Phillips-etal-1999}, \citet[S99]{Shahbaz-etal-1999}, \citet[GBO01]{Greene-etal-2001}, \citet[BP02]{Beer-Podsiadlowski-2002} and \citet[S03]{Shahbaz-2003}. The upper part of the table gives the values extracted from the considered publications. See the notes below for additional information. The lower part of the table provide calculated quantities where the errors are estimated by computing upper and lower allowed ranges. See text for details. The values of mass ratio, inclination angle and masses of \citet{Orosz-Bailyn-1997} are taken from their Table~7, using $3\sigma$ values. The computed distance is calculated with equations~\ref{equ_magdist}, \ref{equ_ebmv}, \ref{equ_AV} and \ref{equ_R}, the above values of the color excess and absolute magnitudes, and using the color of the comparison star of \citet{Foellmi-etal-2006b}: $(B-V)_0 = (B-V)_{\textrm{\scriptsize HD156098}} = 0.5$, consistent with the results of \citet{Beer-Podsiadlowski-2002}. Despite the wealth of data, its variety, and the even larger wealth of calculations, it is hard to determine which study actually provides accurate results. {\bf Notes}: $^{\mathbf{a}}$ Observations made during outburst. The visual magnitude of the source varied by 0.9 mag during the observations. $^{\mathbf{b}}$ The reference given for the orbital period is \citet{Orosz-Bailyn-1997}, but the actual value quoted is that of \citet{vanderHooft-etal-1998}. $^{\mathbf{c}}$ The authors mention a range from 192 to 214~km~s$^{-1}$ assuming $q=3.0$. $^{\mathbf{d}}$ The authors also mention a mass ratio range $q = 2.93-4.20$ for a flaring angle of the disk $\gamma = 10^{\circ}$ instead of the adopted $\gamma = 2^{\circ}$. $^{\mathbf{e}}$ Actually, the author uses the value of $i = 70.2\pm1.6^{\circ}$ from \citet{Greene-etal-2001} when computing the component masses, although the value from \citet{Orosz-Bailyn-1997} is mentioned in the introduction. $^{\mathbf{f}}$ The luminosity is considered fixed, since the mass function is fixed and its value is taken from \citet{Bailyn-etal-1995b}. However, the authors also test their model with $L$ = 31 and 51 $L_{\odot}$. We also note that the authors use the symbol $L_{\textrm{\scriptsize opt}}$ to designate a bolometric luminosity. $^{\mathbf{g}}$ Polar temperature of the star. $^{\mathbf{h}}$ The temperature is adopted from an F6III spectral type determined by \citet{Shahbaz-etal-1999} following the temperature tables of \citet{Gray-1992}. $^{\mathbf{i}}$ The value of the color excess (that is used to compute also the observed luminosity), is taken from the IAU Circular by \citet[][see also Appendix~\ref{graph}]{Horne-etal-1996}. $^{\mathbf{j}}$ This velocity is computed with equation~\ref{equ_vsini}. In the article the authors refer to the velocity determined by \citet{Israelian-etal-1999}: $v \sin i = 93\pm3$~\kms. $^{\mathbf{k}}$ Assuming here $T_{\textrm{\scriptsize eff}} = 6500$K as in \citet{Orosz-Bailyn-1997}. $^{\mathbf{l}}$ Assuming here our derived absorption $A_V = 3.49$, and the apparent magnitude of \citet{Orosz-Bailyn-1997}: $V = 17.2$, which is equivalent to the mean $V$ magnitude determined by \citet{Greene-etal-2001}.}
	\label{orbital-parameters} 
\end{sidewaystable*}

\subsection{Notes on individual studies.}

The paper by \citet{Bailyn-etal-1995b} is not discussed because it uses observations that are the first on this target in the optical domain, and as such, provides only very few quantitative results. We also ignore the study by \citet{Hynes-etal-1998}, who present a large dataset, and discuss extensively the problem of the extinction towards \gro, but use basically all dynamical parameters from \citet{Orosz-Bailyn-1997}. Other studies, such as \citet{Buxton-Vennes-2001} or \citet{GonzalezHernandez-etal-2008} provide updates to a small subset of the orbital parameters, but either have large uncertainties or refer heavily to values determined elsewhere. We think that none of the studies mentioned in this section obtained an accurate and complete parameter set about \gro. It is nevertheless insightful to excerpt specific issues that compromise the accuracy of the quantitative results.

\subsubsection{\citet{vanderHooft-etal-1997}}

The main issue in \citet{vanderHooft-etal-1997} is the fact that observations were taken during the outburst, and the overall magnitude increases by 0.9 mag during the course of the observations. Their folded $R$-band lightcurve clearly shows two different levels of maximum (see their Fig.~3). The authors finally determine the mass ratio by fitting  their lightcurve with a basic eclipsing-disk model with X-ray heating and two estimations of the X-ray luminosity communicated privately by Harmon. Given the intrinsic variability seen in the data, it is clear that only estimations can be made. With the component masses and the orbital period, it is in principle possible to derive a value for $K_2$ and calculate the rotational velocity. However, given the large uncertainties on the mass ratio and the secondary mass, it would be meaningless. We nonetheless note that, if the system is synchronous, the allowed range for the equatorial Roche velocity (70.5--86.2~\kms) is in disagreement with a projected (and hence minimum) rotational velocity of 93$\pm$3\kms determined spectroscopically by \citet{Israelian-etal-1999} and with which \citet{Foellmi-etal-2006b} totally agree.

\subsubsection{\citet{Orosz-Bailyn-1997}}

\citet{Orosz-Bailyn-1997} used optical spectra taken during the outburst (in 1995) and outside the outburst (in 1996), as well as $BVRI$ photometric lightcurves (only in 1996). Their spectroscopic data comprises the data published in \citet{Bailyn-etal-1995b}. Despite the fact that the source was variable and the secondary star is filling its Roche lobe, they fit a sine wave to their radial velocity data (see their Fig.~1). This issue is corrected later by \citet[][but see below]{Phillips-etal-1999}. In order to determine the inclination and the mass ratio, they fit their lightcurves with a model comprising 16 parameters, among which 9 are fixed, and where the spectrum of the star is approximated by a blackbody. This latter issue is corrected in \citet[][also discussed below]{Beer-Podsiadlowski-2002}. 

Concerning the distance, if we take the observed value of the apparent magnitude $V = 17.12$, their luminosity $L = 46.6 L_{\odot}$, combined with $E(B-V) = 1.3$ and a standard $R=3.1$, we obtain $D = 3.1$~kpc. However, if we take the Roche luminosity (hence assuming synchronicity, as did the authors) inferred from their precise mass $m_2$, the orbital period $P$, and the visual absorption value we determined earlier ($A_V = 3.49$), we obtain $D = 3.8\pm0.2$, i.e. larger than the upper limit determined by \citet{Hjellming-Rupen-1995} with the radio jets. Moreover, we note that the value of the luminosity decreases by a factor of two in the corrected model of \citet[][see below]{Beer-Podsiadlowski-2002}.

\subsubsection{\citet{vanderHooft-etal-1998}}

\citet{vanderHooft-etal-1998} attempted to fit their $VRi$ lightcurves with an ellipsoidal model in which the secondary mass $m_2$ and the inclination angle $i$ are left free. However, they used a fixed mass function that has been determined with the early data taken by \citet{Bailyn-etal-1995b} during outburst. Once the mass function is known, the primary mass, and thus the mass ratio (required for the model) are known. Moreover, they assume a bolometric luminosity (strangely called $L_{\textrm{\scriptsize opt}}$ in the article) $L = 41 L_{\odot}$ taken from \citet{Orosz-Bailyn-1997}, which does not corresponds to the correct value: $L = 46.6 L_{\odot}$. They also assume $E(B-V) = 1.3$ mag taken from the IAU Circular by \citet{Horne-etal-1996} discussed in Sec.~\ref{sec_hst_stis}. With such fixed values determined elsewhere from outburst data, it is unlikely that the outcome will be accurate.

In Table 3, we note that the distance implied by their study on the distance cover the range 3.0--4.2~kpc, due to the fact that they have a very uncertain $m_2$, hence mass ratio, combined with a poorly constrained inclination angle.

\subsubsection{\citet{Phillips-etal-1999}}

\citet{Phillips-etal-1999} reanalyse the outburst spectroscopic data only of \citet{Orosz-Bailyn-1997}, ignoring the photometric datasets. They show that X-ray heating of the secondary surface can significantly modify the radial velocity curve. They finally obtain a semi-amplitude of $K_2 = 196$~\kms, i.e. much lower than the previous value, and therefore derive an "updated" value of the mass ratio. Using the the allowed range for inclination angle determined by \citet{vanderHooft-etal-1998}, they can compute a new mass for the black hole. This study certainly demonstrates the need to correctly take into account the X-ray irradiation when modelling the radial-velocity curves \citep[or alternatively, re-emphasize the need to choose a truly quiescent optical lightcurve; see again][]{Cantrell-etal-2008}. But it does not provide accurate results for the \gro\ system itself.

We note in Table~\ref{orbital-parameters} that their results imply a smaller rotational velocity $v \sin i = 75\pm5$~\kms, which is again in large disagreement with the measured value mentioned above. 

\subsubsection{\citet{Greene-etal-2001}}

\citet{Greene-etal-2001} obtained photometric lightcurves in the optical ($B$, $V$ and $I$, said to be "quantitatively indistinguishable from that of \citet{Orosz-Bailyn-1997} and \citet{vanderHooft-etal-1998}"). They also obtained for the first time near infrared data (in $J$ and $K$ bands) in order to have better constraints on the role of the accretion disk. Moreover, their modelling is based on the code presented in \citet{Orosz-Hauschildt-2000} which is more sophisticated than what Orosz \& Bailyn used. The authors mention that they make use of "all of the binary system observables and their uncertainties", mentioning the 29 points of the radial velocity curve of \citet{Shahbaz-etal-1999} as their second dataset (the first one being their own photometry, and the third being the rotational velocity of \citet{Israelian-etal-1999}). Table~\ref{orbital-parameters} illustrates clearly that the results of \citet{Greene-etal-2001} are in perfect agreement with that of \citet{Shahbaz-etal-1999}, but not with that of \citet{Beer-Podsiadlowski-2002} who use basically identical photometric data (see below).

The reason of the discrepancy is provided by \citet{Beer-Podsiadlowski-2002} who discuss extensively the results of \citet{Greene-etal-2001}. Similarly to the work of \citet{Orosz-Bailyn-1997}, Greene and coworkers implicitly allowed for arbitrary offsets between the different lightcurves, which are not fitted simultaneously (in addition to not using any extinction and distance information). Greene et al. concluded that the accretion disk does not contribute to the lightcurves. On the other hand, \citet{Beer-Podsiadlowski-2002} found a much poorer fit of the multi-color data when no disk is present, concluding that no fully self-consistent models can be obtained without a disk.

\subsubsection{\citet{Shahbaz-etal-1999} and \citet{Shahbaz-2003}}

\citet{Shahbaz-etal-1999} analyse new optical spectra taken during X-ray quiescence and find $K_2 = 215$~\kms. The data was then reanalyzed by \citet{Shahbaz-2003} who developed a procedure to determine the spectroscopic mass ratio in interacting binaries only by modelling the secondary star spectrum only.

There are various issues with this approach, but the main one is with the data. In the introduction, the author emphasizes the numerous problems associated with the developed procedure (intermediate-resolution spectroscopy only, shape of the rotational profile, wavelength-dependent limb-darkening, Roche-lobe filling and so on), but the solution proposed to reduce "these uncertainties" is to "determine the exact rotationally broadened spectrum from the secondary star in an interacting binary", i.e. make the best possible model. It remains true however that even a perfect model cannot work well with medium quality data.

As matter of fact, the spectral range is comprised in a very small region between 6320\AA\ and 6550\AA, where there are many iron lines, with a resolution of 4.2\AA. With an observed rotational velocity of 93~\kms\ (see above), the line are heavily blended. However, an intermediate step of the procedure involves fitting the continuum of the spectrum. This issue is crucial, since the procedure rely on the strength of the lines, which obviously depends critically on the true continuum level and the amount of possible veiling by the accretion disk. There is no detailed indication about this central step, apart that a spline fit is made before combining the spectra (which is also not an easy task). This issue is already present in the analysis of the abundances in \gro\ secondary star by \citet{Israelian-etal-1999} and which has been questioned in \citet{Foellmi-etal-2007a} who demonstrate the impossibility to measure abundances as precisely as it has been claimed\footnote{The issue has not been truly addressed in \citet{GonzalezHernandez-etal-2008}, where it is still not demonstrated how the continuum location is determined.}. Even with a Signal-to-Noise ratio of 100, it is very difficult to know where is the continuum in the spectrum, if any is truly present given the very broad lines. This requires a spectrophotometric calibration of excellent quality (one or two \% maximum), which is the case neither in \citet{Shahbaz-etal-1999} nor in \citet{Israelian-etal-1999}.

Interestingly, \citet{Shahbaz-2003} obtain a poor first fit that they attribute partially to the fact that some elements might be overabundant as it has been supposedly shown by \citet{Israelian-etal-1999}. Given the amount of uncertainties and the quality of the data used, we can safely discard the results of this study. Moreover, the distance implied by the results is the largest of the height studies compared in Table~\ref{orbital-parameters} ($D = 4.0\pm0.5$).

\subsubsection{\citet{Beer-Podsiadlowski-2002}}

\citet{Beer-Podsiadlowski-2002} reanalyze the photometric data of \citet{Orosz-Bailyn-1997}, and ignore their spectra. They developed a better lightcurve modelling by not assuming a perfect blackbody spectrum for the secondary star, which has strong consequences on the relative fluxes between the $B$ and the $V$ filter bandpasses. Moreover, they fit {\it simultaneously} in a self-consistent manner the lightcurves in the different passbands. One can see in Table~\ref{orbital-parameters} that their results imply a decrease of luminosity by a factor $\sim$2, and a smaller mass of the secondary star by 40\% compared to that of Orosz \& Bailyn. 

The problem in the models of \citet{Beer-Podsiadlowski-2002} is that the distance is said to be a free parameter, although it is not clear if it has been allowed to go as low as 1.0 kpc. In many places the authors claim that the distance of 3.2 kpc of \citet{Hjellming-Rupen-1995} has been used "to tighten" their results, since the three main parameters (distance, color excess and temperature) are highly correlated, and can compensate for each other. But if $D$ is basically fixed, what room is left for the other parameters? Interestingly, \citet{Beer-Podsiadlowski-2002} found it reasonable to use a distance of 3.2~kpc since the distance by \citet{Hjellming-Rupen-1995} is consistent with the other determination of \citet{Bailyn-etal-1995a}, \citet{Greiner-etal-1995}, \citet{McKay-Kesteven-1994} and \citet{Tingay-etal-1995}, although they also note that the existence of a significant difference between the jet inclination ($\sim85^{\circ}$) and the disk inclination angles makes the model of the wiggles in radio jet data not necessarily appropriate. We have shown above how problematic are these references.

Let us mention here that a distance of 1.0~kpc implies a velocity $\beta = 0.28$ and consequently an inclination angle of the jets of $\sim 71^{\circ}$. It means that at 1.0~kpc, \gro\ would probably not have its disk misaligned with the jets to the contrary of what has been claimed \citep[e.g.][]{Maccarone-2002}.

Finally, we need to emphasize here that the first best-fit steady-state disk model in \citet{Beer-Podsiadlowski-2002} is discarded by the authors because it was giving a distance "much lower than 3.2~kpc". The work by \citet{Beer-Podsiadlowski-2002} is certainly the best multi-color modelling of the lightcurves of \gro\ to date. Unfortunately, it remains unknown what the solution of their model was with a much smaller distance.

\subsection{Summary}

Despite the wealth of data, its variety and the even larger wealth of calculations made either on the data itself, or on the interpretation of the quantitative results (evolutionary tracks, Monte Carlo simulations, $\chi^2$ distance minimization between spectra, etc.), it is hard to determine which study actually provide accurate, and not only precise, results. It is partly due to the fact that each paper studies only a subset of all parameters, taking the missing ones from other publications. 

To summarize, we have clearly demonstrated how difficult the modelling of an interacting binary such as \gro\ is, and that a complete and consistent modelling of the photometric {\it and} spectroscopic datasets taken during true {\it optical quiescence} is still lacking.

\section{A new estimation of the distance of \gro}

\begin{figure}
	\centering
	\includegraphics[width=0.8\linewidth]{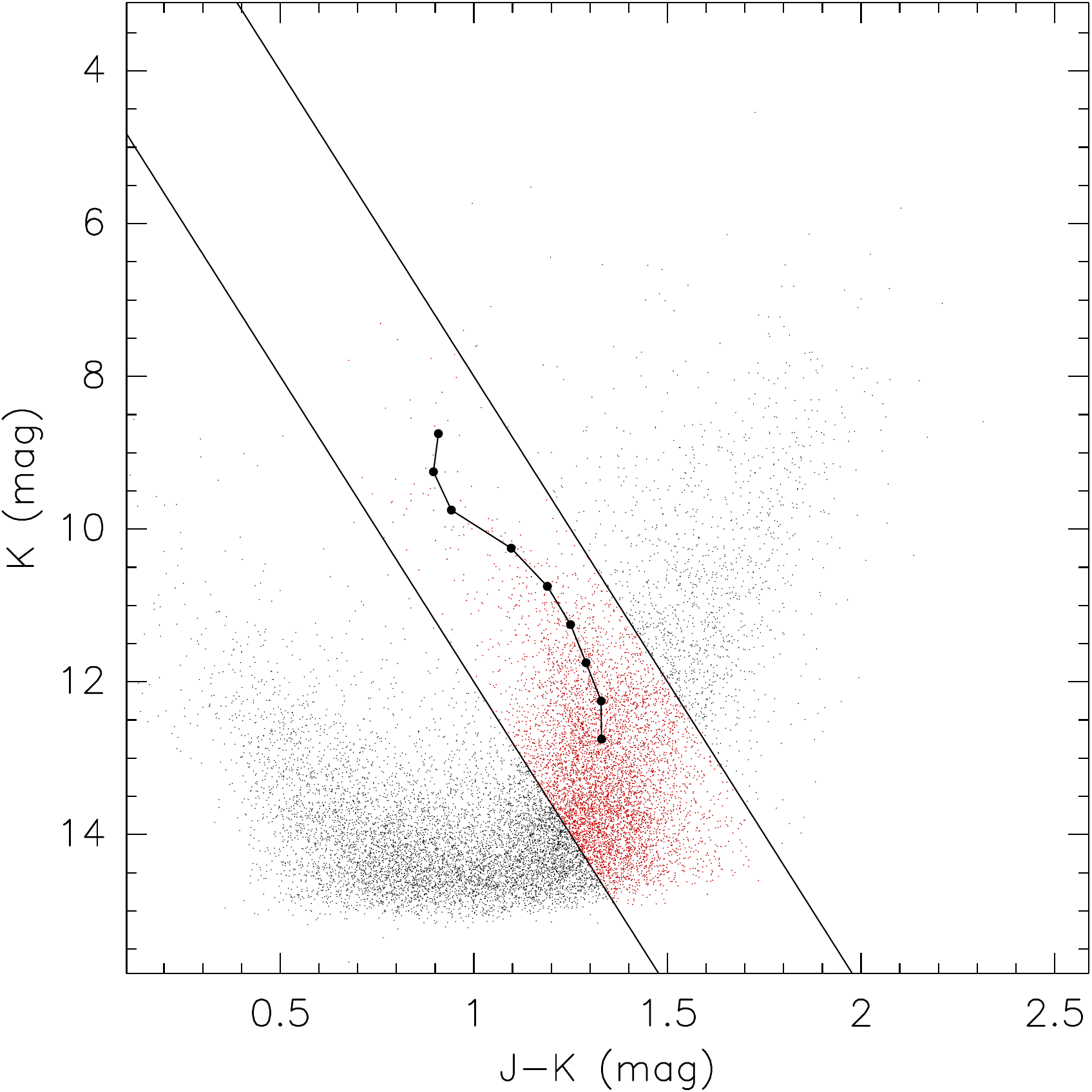}
	\caption{Color-Magnitude diagram of the stars in a square of $0.45^{\circ}\times0.45^{\circ}$ around the position of \gro\ extracted from the 2MASS database. The two straight lines delimitate the selected region (red dots). The linked points indicate the maximum number density of selected stars in every magnitude bins. By using a larger area, we increase the number of stars and thus the contamination of the sample. On the other hand, using a smaller area make the results less precise.}
	\label{fig_2mass_k_j-k}
\end{figure}

\begin{figure}
	\centering
	\includegraphics[width=0.8\linewidth]{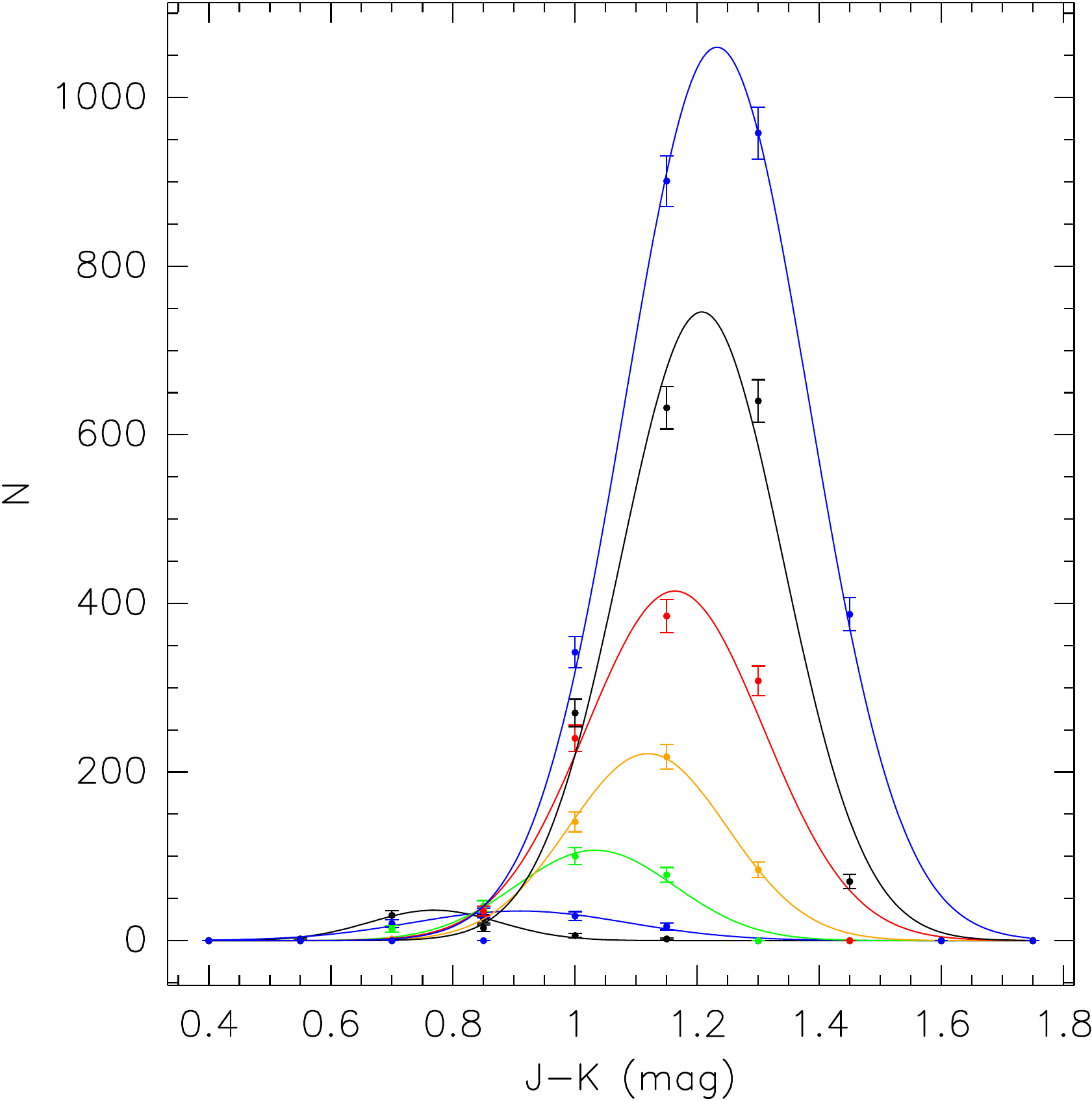}
	\caption{Gaussian fits to the histograms of the number of selected stars for each magnitude bins as a function of the $J-K$ color. One can see that there is a progression of the maximum of the gaussian peaks towards redder colors when we go from bright stars (lower curves with peaks to the left) to fainter stars (upper curves).}
	\label{fig_2mass_gaussians}
\end{figure}

Recently, \citet{Guver-etal-2008} have derived the distance of the galactic neutron star 4U~1618-52 using the fact that red clump giant stars can be considered as standard candles \citep[see][where the method is well explained]{LopezCorredoira-etal-2002}. Following this method, we retrieved 2MASS {\it JHK} magnitudes of stars in a box of $0.45^{\circ}\times0.45^{\circ}$ around the position of \gro. On the 14659 stars, we constructed a Color-Magnitude diagram (CMD) and selected the red clumps giants in a similar way to what is done in \citet{Guver-etal-2008}, and as shown in Fig.~\ref{fig_2mass_k_j-k}. We obtained 6033 stars that we binned in magnitudes with bins of widths 0.5 mags. For each bin, we fitted a gaussian on the histogram of the number of stars as a function of the color $J-K$. Each gaussian fitted peak provides a value of the color $J-K$, which is then identified to the color of red clump giant Stars of the given magnitude bin (Fig.~\ref{fig_2mass_gaussians}). Identically to \citet{Guver-etal-2008}, we then assume an absolute magnitude of $K = -1.62$ and a unabsorbed color $(J-K)_0 = 0.7$ for these stars. \citep[See also][for slightly different values and a little variant to this method.]{Durant-vanKerkwijk-2006}

\begin{figure}
	\centering
	\includegraphics[width=0.8\linewidth]{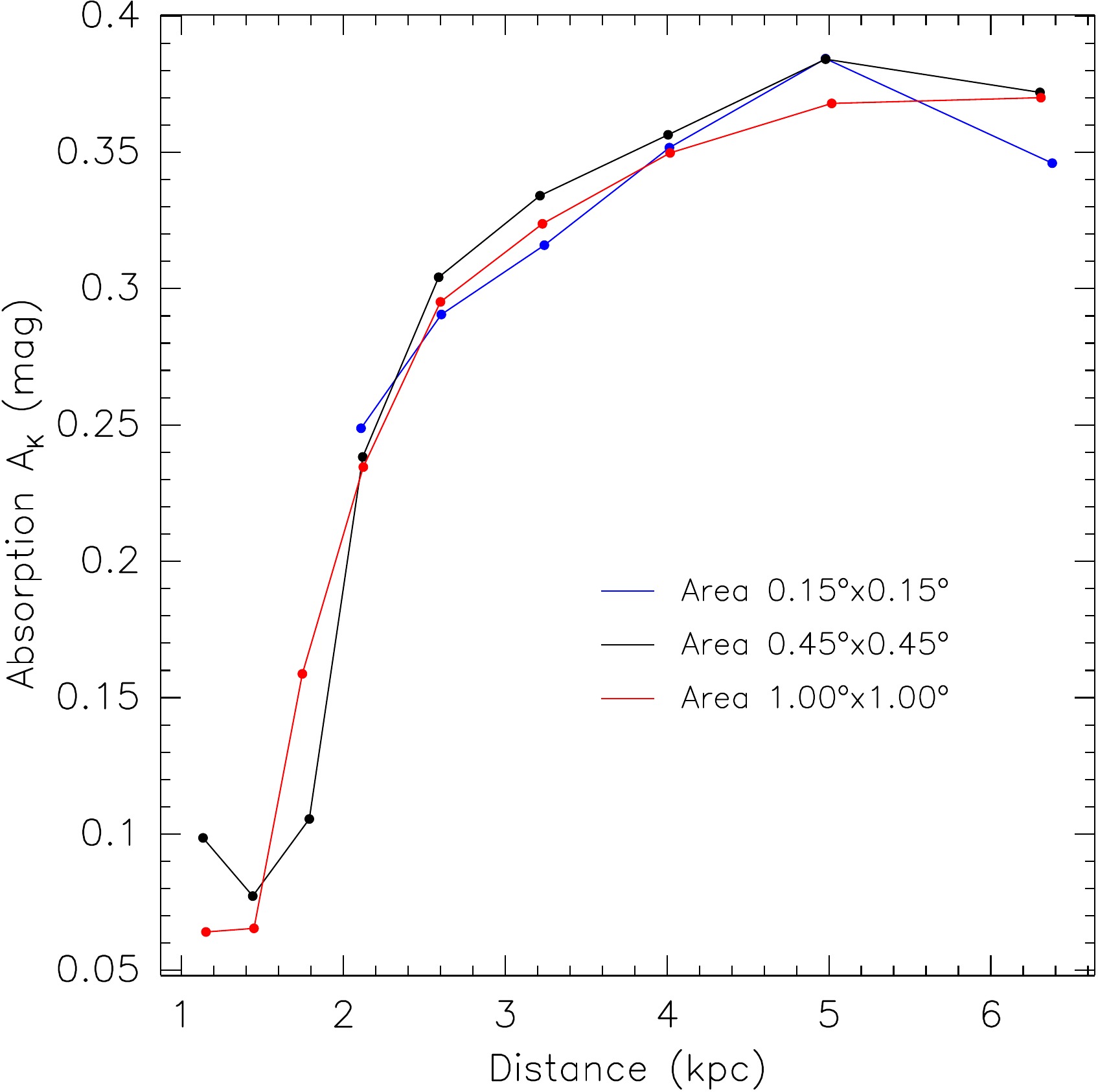}
	\caption{Calculated relationships between the absorption $A_K$ and the distance using the selected stars of the three different areas centered around \gro. The curves saturate around $D \sim 5$ kpc, or $A_K \sim 0.4$, certainly because of the increased amount of contamination of dwarfs stars and M-giants, as mentioned by \citet{Guver-etal-2008} (non-completeness of the 2MASS data for faint magnitudes might also play a role, for the faintest ones). Similarly, at small distance, there are too few stars to obtain a perfectly reliable curve.}
	\label{fig_Ak_D}
\end{figure}

By using the general equation~\ref{equ_magdist}, the following relationship between the absorption magnitude in the K-band and the color indices:
\begin{equation}
\label{equ_Ak}
A_K = c_e \times [(J-K) - (J-K)_0]
\end{equation}
with $c_e = 0.657$ (see Guver et al.), we can derive a relationship between the distance and the absorption derived from the color, in this direction of the sky (see Fig.~\ref{fig_Ak_D}). We have also derived this relation with stars included in areas of $0.15^{\circ}\times0.15^{\circ}$ and $1.0^{\circ}\times1.0^{\circ}$ around the position of \gro (70527 stars in total and 25108 stars selected for the box of one degree squared). It does not change the result significantly as shown in Fig.~\ref{fig_Ak_D}. We also made tests on the extent of the selected region in the CMD, and by varying the number, sizes and positions of the magnitude bins, but again it proved to have a negligible influence on the final curves.

The problem is estimating the $(J-K)$ and $(J-K)_0$ colours in equation~\ref{equ_Ak} for \gro. The former can be visually estimated from the $J$ and $K$ lightcurves by \citet{Greene-etal-2001}: $J-K \sim 0.55$ \citep[][use $J-K = 0.6$]{Beer-Podsiadlowski-2002}. As for the later, we need to assume that the infrared color of the secondary star in \gro\ can be compared to that of an unaffected star of similar spectral type (F6IV). \citet{Koornneef-1983} gave $J-K$ ranging from 0.24 to 0.29 for F2 to F8 {\it dwarf} stars\footnote{No values are given for F giants, but $J-K$ ranges from 0.22 to 0.35 for F2 to F8 supergiants.}. On the other, one can also use the 2MASS values of the comparison star HD~156098 of \citet{Foellmi-etal-2006b} which is 50 pc away from the sun and can be considered as unaffected by extinction. Its infrared color is $(J-K ) \equiv (J-K)_0 = 0.397$. Using the extreme values of the given range $0.24 < (J-K)_0 < 0.4$, and equation~\ref{equ_Ak}, we derive a $K$-band absorption $0.1 < A_K < 0.2$ for \gro. A comparison with the curves of Fig.~\ref{fig_Ak_D} shows that the distance of \gro\ is certainly less than 2.0~kpc. 

We note however that it is still not possible to definitely conclude the exact distance of \gro\ (i.e. it lies at the same distance as the cluster NGC~6242) because of the impossibility to derive a reliable curve for the brightest (and thus nearby) red clump giant stars. However, these new results confirm that of Foellmi et al.: \gro\ is likely to be much closer than currently admitted.

\section{The distance of \aaa}

We have presented above a confirmation that the distance of \gro\ is certainly smaller than 2~kpc. One can thus ask: is \gro\ the closest (stellar) known black hole to the Sun? According to \citet{Jonker-Nelemans-2004}, \aaa\ is at a distance of 1.2$\pm$0.4~kpc from the Sun and is so far the closest known black hole to the Sun\footnote{We obviously ignore here the possible problems with the determinations of all other microquasars listed in \citet{Jonker-Nelemans-2004}, and in particular XTE~J1118+480 which has a published distance of $D = 1.8\pm 0.6$~kpc.}. However, its distance is also problematic, but for different reasons than that of \gro. We start by critically review the published distance of \aaa\ before presenting a new estimation of its maximum distance.

\subsection{What is the the color excess towards \aaa?}

In their study of distances of low-mass X-ray binaries, \citet{Jonker-Nelemans-2004} quote \citet{Shahbaz-etal-1994} and \citet{Barret-etal-1996} for the distance of \aaa. Shahbaz and collaborators give a distance range between 650 and 1450~pc, with a preferred value of 1050~pc. 

To estimate the extinction toward \aaa\, \citet{Shahbaz-etal-1994} use an estimation of the color excess $E(B-V) = 0.35$ from \citet{Wu-etal-1983}. But Wu and collaborators actually quote their own results obtained a few years earlier: \citet{Wu-etal-1976}. This latter paper describe that the extinction is measured by filling the "extinction spectral feature" at 2200\AA\ in their UV spectrum. However, the spectrum of \citet{Wu-etal-1976} has been obtained with the {\it Astronomical Netherlands Satellite}, and consists of no more than 5 points only, respectively at the central wavelengths of 1550, 1800, 2200, 2500, and 3300\AA, simply because the instrument onboard the satellite had only 5 channels. One can question the reliability of such measurement given the extremely low resolution, and the imperfect fit used to derive the color excess.

Interestingly, \citet{McClintock-Remillard-2000} published a \emph{HST/STIS} spectrum of \aaa\ ranging from 1900 to 3100 \AA, clearly revealing the {\it absence} of a "feature" at 2200\AA\ (see their Fig.~1, upper panel). This absence of any strong extinction feature in the UV spectrum of \citet{McClintock-Remillard-2000} \citep[who cite ][ for the distance]{Barret-etal-1996} seems to show that the extinction might be particularly low. This is consistent with a location of \aaa\ far from the Galactic plane (l=209.96$^\circ$, b=6.54$^\circ$; R.A.=6$^h$22$^m$44.4$^s$, Dec.=$-00^\circ$20$^m$45$^s$). 

More recently, \citet{Gelino-etal-2001b} determine a distance of 1164$\pm$114~pc for \aaa\ using the color excess of \citet{Wu-etal-1976}. This distance has also been used by \citet{Shahbaz-etal-2004}. \citet{Gallo-etal-2006} still use the older value of 1.2$\pm$0.4~kpc citing \citet{Shahbaz-etal-1994}, \citet{Gelino-etal-2001b} and \citet{Jonker-Nelemans-2004}. Finally, we note that \citet{Pal-Chakrabarti-2005} still use the distance range of \citet{Shahbaz-etal-1994}, and that \citet[][who obtained $D = 1.4$~kpc]{Esin-etal-2000} cite \citet{Shahbaz-etal-1994} and \citet{Barret-etal-1996}. All these studies, even those being very recent, directly rely on the 30-years old determination of $E(B-V)$ by \citet{Wu-etal-1976}, even if they do not cite the original paper.

\subsection{A wealth of uncertain estimations}

\citet{Barret-etal-1996} use two different methods to estimate the distance of \aaa, and found a value (1.2~kpc) "in agreement with previous determinations". We note that they do not apply either method to \gro\ (which is also discussed in the paper) for which they take the literature value, 3.2 kpc.

The first method is the magnitude-comparison method (equation~\ref{equ_magdist}), assuming that the Roche-lobe radius derived from dynamical studies is equal to the "radius" of the secondary star. Moreover, Barret and coworkers state that the absolute magnitude depends only on the spectral type. Using another method to compute the Roche radius requiring a mass assumption ($M = 0.4 M_{\odot}$ for mid-K secondaries), Barret and collaborators confirm their distance value: $D = 1.2$~kpc. No uncertainties are provided, but they estimate them at about 25\%. They also cite \citet{Haswell-etal-1993} and \citet{Shahbaz-etal-1994}, who provide the optical derredened magnitude and hence rely on the non-existent feature measured by \citet{Wu-etal-1976}. 

\citet{Oke-Greenstein-1977}, cited by \citet{Jonker-Nelemans-2004} on their Table~3. discussed in Sec.~\ref{sec_X-rays}, estimate the color excess using simultaneously the relations of \citet[][$E(B-V) = 0.44$ with a large uncertainty]{Spitzer-1948}, \citet[][$E(B-V)$ ranging from 0.25 to 0.6]{Wampler-1966} and \citet[][]{Aannestad-etal-1973}. But Spitzer use the data of \citet{Stebbins-etal-1940}, and Aannestad and coworkers use that of \citet{York-1971}. \citet{Oke-Greenstein-1977} conclude that "these methods cannot rule out much larger distances, since the object is largely out of the galactic plane". Finally, they use a distance of 870 pc using another estimation of the absolute magnitude of the secondary dwarf star, and claim that this new value is "somewhat smaller than previous estimates [...] but not inconsistent with them".

To summarize, it appears that all methods are more or less giving about the same result around 1~kpc, but with still a rather large scatter around this value. We also note that most estimations rely on the determination of the color excess by \citet{Wu-etal-1976}, even in most recent literature. 

Below, we use VLT/UVES spectra to apply the maximal-distance method of \citet{Foellmi-etal-2006b} which, even systematically uncertain, provides an model-independent estimation of the distance. Interestingly, we can measure the equivalent width of the sodium doublet in these UVES spectra of \aaa: $EW = 0.5\pm0.1\AA$. The value is slightly outside the allowed range given in Sec.~\ref{sec_sodium}, and more importantly,   equation~\ref{equ_dellaValleDuerbeck} might simply not be applicable to an object outside the galactic plane, in the direction of the anticenter, since it is outside the directions where the relationship has been calibrated. If we use it nonetheless with caution, we obtain $E(B-V) = 0.22\pm 0.07$. Assuming $R=3.1$, we have $A_V = 0.68$. According to \citet{Shahbaz-etal-1994}, the secondary star radius is 0.8 $R_{\odot}$, its temperature 4000~$K$ (implying a luminosity of $L = 0.14 L_{\odot}$, that is an absolute magnitude of $M_V = 7.0$), its visual magnitude $V = 18.2$ and 40\% disc contamination (i.e. a magnitude increase of 0.55 mag). The corresponding distance is 1.3~kpc, and 1.6 without veiling. 

Could actually \aaa\ be further away? Given the various uncertainties, especially on the secondary star radius, this distance is certainly not conclusive.

\subsection{Astrometric estimations of the proper motion}

We looked in public archives for images that could provide an astrometrical estimation of the distance of \aaa. Unfortunately, it has not been observed with an imaging technique by \emph{HST}. We were finally  able to find only acquisition images in the ESO archive with the following instruments: 3.6m/EFOSC2 (ESO, La Silla Observatory), VLT/ISAAC and VLT/FORS1 (ESO, Cerro Paranal Observatory). Unfortunately, ISAAC frames proved to be useless because of the too small number of stars in the field of view, and were discarded. Finally, we also retrieved SuperCOSMOS images, which have a much poorer pixel scale but provide the largest baseline in time. Table~\ref{tab_astrometry} summarize the properties of the three datasets.

\begin{table*}
	\small
	\centering
	\caption{Summary of the archival data used to derive astrometry of \aaa. The instrument/project, observatory location, date of observations and bandpass (filters) are indicated as well as the Field-of-View in arcminutes, and the pixel scale (in arcseconds per pixel). The pixel scale of SuperCOSMOS plates are computed given a plate scale of 67.14 '/mm and 10$\mu m$ pixel size. The FOV of SuperCOSMOS images is originally 6.4x6.4 degrees, but smaller regions can be retrieved electronically.}
	\begin{tabular}{ccccccc} \hline 
	Instrument  & Location & Date & Filter (name) & FOV ('x') & Scale (''/pixel) \\ \hline
	SuperCOSMOS/UK Schmidt & Siding Spring Obs. & 16/02/1983  & I (RG715) & up to 384x384 & 0.6714 \\
	VLT/FORS1   & VLT/Paranal     & 7/01/2003  & V (V\_BESS+35) & 6.8x6.8  & 0.200  \\
	La Silla/EFOSC2  & La Silla (3.6m) & 16--18/01/2005 & V (V\#641)     & 5.3x5.3 & 0.31 (2x2 bin) \\ 
	\hline
	\end{tabular}
	\label{tab_astrometry}
\end{table*}

In order to check if the star has been ejected from a cluster with a runaway velocity, similarly to GRO J1655-40, we performed astrometrical calculations with the above archives images. We chose the EFOSC2 image to be the reference since it had the sharpest PSF, and selected 40 bright isolated stars that were visible on all other frames. We computed the geometrical transformation map between the reference and the other images using the fitted positions of these 40 stars. Finally, we computed the expected position of the target on the EFOSC2 image for a given transformation matrix and its actual position on this image. The difference in pixels was transformed into arcsecond using the pixel scale of the EFOSC2 CCD. Finally, the results were divided by the time elapsed between the reference and the given image. 

It appeared that the best transformation fit was obtained between the FORS1 and EFOSC2 images, with an rms = 0.1508 and 0.1506 pixels in the X and Y direction respectively. This translates into an uncertainty of about 30 mas. The very small observed difference of the position of \aaa\ between the FORS1 and EFOSC2 images was clearly below this value: -10.4 and 9.0 mas in the X and Y directions respectively. For the SuperCOSMOS images, the rms achieved in the transformation fit is 0.65 and 0.20 pixels, corresponding to a uncertainty of 436 and 134 mas respectively. The observed displacement was again inside the uncertainties of 71 and 136 mas.

Taking the 2.03 years difference between the FORS1 and EFOSC2 images, we obtain a maximum proper motion of about 15-20 mas/yr, which roughly corresponds to that obtained with SuperCOSMOS images that are 22.08 years apart from EFOSC2 ones. It is possible to look at the possible presence of a cluster of stars within a given radius, taking into account this upper limit of the astrometric motion of \aaa\ on the sky. Looking at a region of radius of about 3 degrees using SIMBAD, we found two clusters: [KPR2005] 22 \citep{Kharchenko-etal-2005}, and C~0619+023 \citep{Collinder-1931}. While there is no information on the latter, the former has an estimated distance of 1.5~kpc. It is located 2.2808 degrees away from \aaa, corresponding to a projected separation on the sky of $\sim$59.7~pc at 1.5~kpc. Given the upper limit of the projected proper motion of \aaa, one can estimate its minimal age to be about 367~000 years.

The current data does not allow us to derive strong constraints on the proper motion of \aaa, and therefore on its possible origin. Combined with the systemic radial velocity \citep[4~\kms according to][]{Marsh-etal-1994}, it would be interesting to simulate possible galactic orbits for \aaa, similarly to what \citet{Mirabel-etal-2002} have done for \gro.

\section{The maximal-distance method of Foellmi et al. (2006) and its application to \aaa}

\subsection{Issues in the method}

In \citet{Foellmi-etal-2006b} a new method allowing to estimate a maximum distance has been presented and applied to the case of \gro. It is based on the comparison between the calibrated spectroscopic fluxes of the secondary star and that of a companion star of similar spectral type and luminosity class which needs to be close enough to have a negligible absorption. The comparison is made with spectra obtained with the same instrument (VLT/UVES) configured with an identical setup. The main issue with this method is that one needs to make an hypothesis about the absolute magnitude difference between the two stars. More precisely, if we call $f$ the ratio of the spectroscopic fluxes between the secondary star in the microquasar, and the comparison star, \citet{Foellmi-etal-2006b} derive the following relation:
\begin{equation}
	a = 5 \log \left( \frac{D_2}{D_1} \frac{1}{\sqrt{f}} \right) + M_2 - M_1 \geq 0 
	\label{foellmi} 
\end{equation}
where $a$ is the (spectroscopic) absorption toward the target (i.e. the average absorption within the narrow wavelength range given by the UVES spectra used to make the comparison). $D_1$ and $D_2$ are the distances of the microquasar and the reference star respectively. Obviously, $a$ must be null or positive. The difference between the two absolute magnitudes $M_1$ and $M_2$ is however impossible to evaluate. Or, in other wods, how different is the absolute magnitude $M_1$ compared to that of the reference star, given that the spectra are directly compared. For \gro, \citet{Foellmi-etal-2006b} allowed for a large range and made the hypothesis that $M_1$ is comprised between $M_2 - 1$ and $M_2 + 1$. They finally found that $D \lesssim 1.7$ kpc by comparing flux of \gro\ to that of 4 nearby stars of similar spectral types. However, because the absolute magnitude is unknown, this method has a systematic intrinsic uncertainty. 

Since then, various authors mention this distance revision \citep{Caballero-Garcia-etal-2006-astroph,Takahashi-etal-2008,Joinet-etal-2008}, but simply continue to use the canonical value. \citet{Sala-etal-2007} used \emph{XMM-Newton} data to derive a connection between the inner radius of the accretion disk and the distance. They argue that if the distance is less than 1.7 kpc and its mass is less than 5 $M_{\odot}$, then the inner accretion disk radius will be inside the gravitational radius of the black hole. However, as noted by \citet{Combi-etal-2007}, the mass and spin of the black hole are uncertain. If the black hole were rapidly rotating as suggested by \citet{McClintock-etal-2006} and its mass less than 5 $M_{\odot}$, the horizon of the Kerr black hole could then be well inside the inner accretion radius. Furthermore, \citet{Sala-etal-2007} admit that an estimate of the inner accretion radius using \emph{XMM} data is model dependent. 

\citet{Lasota-2008} mention that if we were to accept the new and smaller distance, the secondary in \gro\ would not be filling its Roche-lobe \citep[a similar remark is made in][who do not provide however any reference about it]{Caballero-Garcia-etal-2007} and in this case too we would have to invoke a mass transfer instability to explain the outbursts. But we have seen above that a complete modelling of the multi-color lightcurves with a good understanding of the disk emission and the absorption is still lacking. 

\subsection{Application to \aaa}

In order to check the distance of \aaa, we applied two different methods. The first one is that of the red clump giant stars. Unfortunately, it proved to be useless because of the too small number of stars in the 2MASS database in the direction of \aaa, even taking a box of one degree around it. We were simply unable to identify a Red Clump Giant star branch in the CMD. It can certainly be explained by the direction of \aaa\ in the sky, which points away from the galactic center. The other method is that of \citet{Foellmi-etal-2006b}. To complement the study, we derive constraints on its proper motion using archival imaging data.

\begin{figure}
	\centering
	\includegraphics[width=1\linewidth]{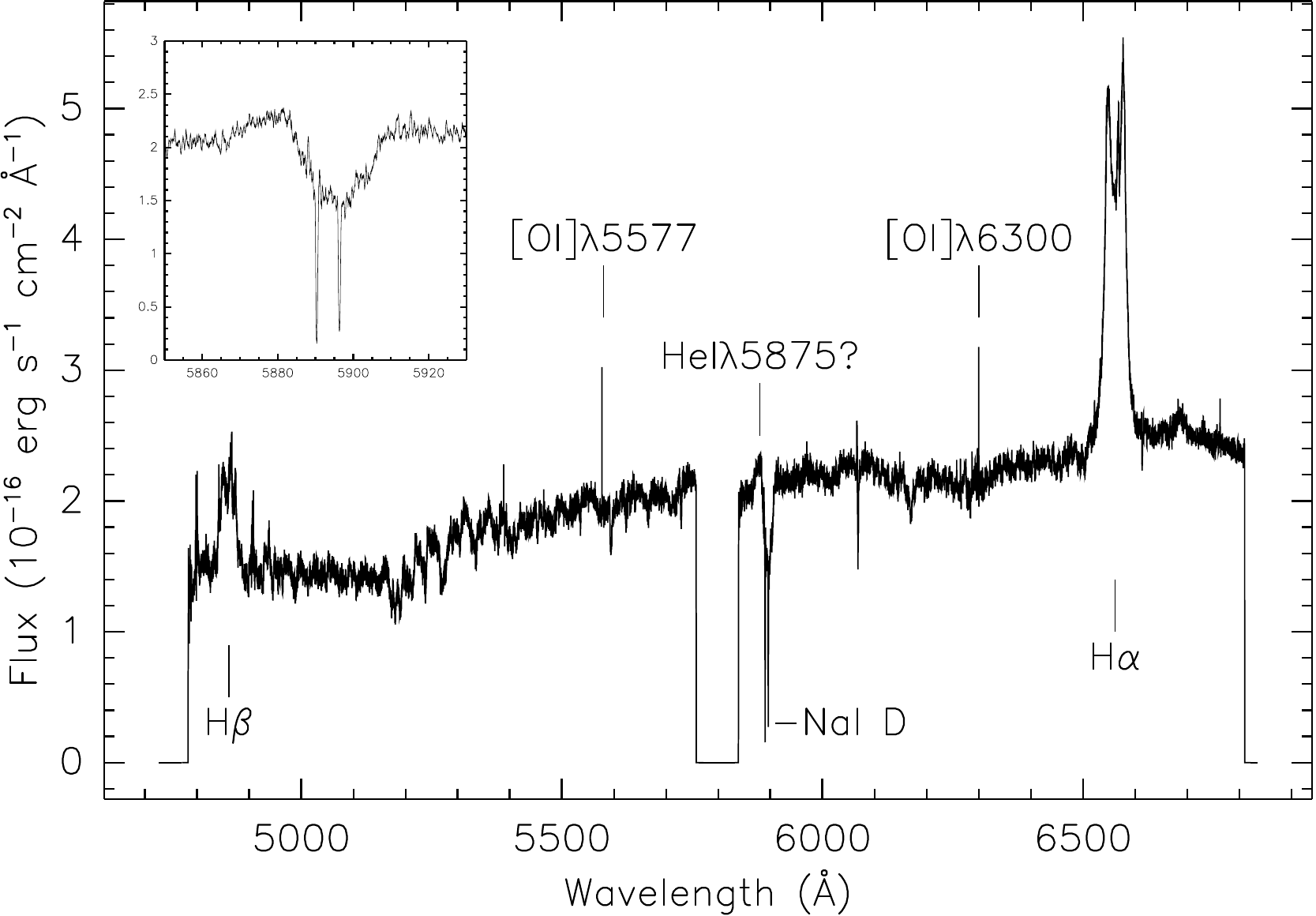}
	\caption{UVES spectrum of \aaa\ (lower and upper CCDs combined), smoothed with a boxcar of 11 pixels. Major spectral features are marked. The inset shows a close-up of the sodium doublet region. \citet{Neilsen-etal-2008} identify with caution the possible presence of HeI$\lambda$5875 in this region.}
	\label{fig_A0620_spectrum}
\end{figure}

We looked for VLT-UVES archival data on \aaa, with the aim at comparing them to the UVES spectra of the  UVESPOP database\footnote{{\tt http://www.sc.eso.org/santiago/uvespop/}} \citep[see][]{Bagnulo-etal-2003}. We have found 20 spectra of the target (ESO program ID 066.D-0157(A), P.I. Maeder), split in two different wavelength range: 4790 to 5755\AA, and 5840 to 6805\AA. The spectra were taken in 2000, December 5 (12 spectra), 17 (2 spectra) and 21 (6 spectra). The have been reduced and calibrated in exactly the same way as for GRO~J1655-40 \citep[see][ to which the reader is referred for a detailed description of the reduction and flux-calibration of the spectra]{Foellmi-etal-2006b}. The average spectrum is shown in Fig.~\ref{fig_A0620_spectrum}. Moreover, we also retrieved from the UVESPOP the spectra of three different single stars with similar spectral types and luminosity classes: HD~10361 (K5V), HD~100623 (K0V) and HD~209100 (K4.5V).

\begin{table*}[!ht] 
	\centering 
	\caption{The three dwarf K stars used to compute the maximum distance of \aaa\ are summarized. The star's name, spectral type, distance (computed from the {\emph HIPPARCOS} parallax), absolute magnitude, ($B-V$) color (obtained from SIMBAD) and the ratio of the flux-calibrated spectrum of \aaa\ with that of the star are indicated. The last column indicates the ranges of maximum distance $D$ of \aaa\ obtained through the constraint of $a \geq 0$ with the two limits of the absolute magnitude $M_1$. See text for details. The quoted $f$ flux ratios are multiplied by a factor 0.6 before estimating the distance to account for a contamination of the accretion disk of 40\% according to \citet{Shahbaz-etal-1994}. The uncertainty on the distance values is $\pm$0.2~kpc. The error on the absolute magnitude $M_2$ is computed from the error on the \emph{HIPPARCOS} distance. } 
	\begin{tabular}{llcclcll} \hline 
	Star&Sp. Type&Distance& $M_2$ & $B-V$ & $f$ & Max. $D_1$ (spec) \\
	& & (pc) & (mag) & (mag) & & (kpc) \\
	\hline HD~10361 & K5V & 6.6$\pm$0.1 & 6.7$\pm$0.2 & 0.85 & $(5.2\pm0.9)\,10^{-4}$ & 0.24--0.59 \\
	HD~100623 & K0V & 9.5$\pm$0.1 & 6.1$\pm$0.1 & 0.81 & $(6.3\pm0.9)\,10^{-4}$ & 0.31--0.77 \\
	HD~209100 & K4.5V & 3.61$\pm$0.05& 6.9$\pm$0.2 & 1.06 & $(3.9\pm0.9)\,10^{-4}$ & 0.15--0.37 \\
		\hline 
	\end{tabular}
	\label{multistar} 
\end{table*}

\begin{figure}
	\centering 
	\includegraphics[width=0.8\linewidth]{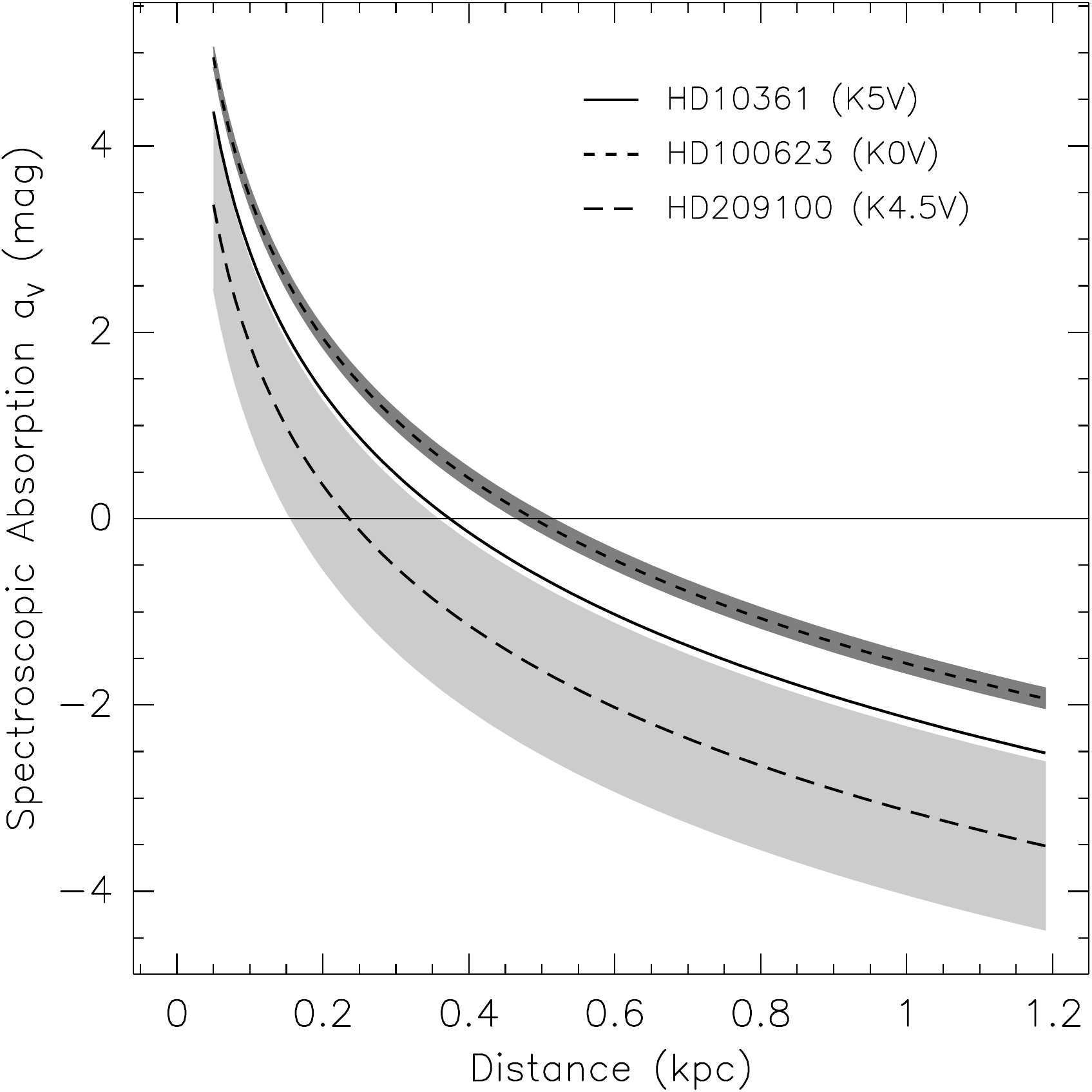} 
	\caption{Relation between the spectroscopic absorption $a$ and the distance, following the equation~\ref{foellmi}. The condition that the absorption must be positive is represented by the horizontal line. The light gray area shows the systematic uncertainty of $\pm$ 1 magnitude, while the dark gray area shows an uncertainty of 10\% on the spectroscopic flux. It shows that the likely distance of \aaa\ is probably smaller than 0.5~kpc.} 
	\label{absD} 
\end{figure}

As in \citet{Foellmi-etal-2006b}, we compare the flux-calibrated spectrum of \aaa\ with that of the comparison stars (ignoring the H$\alpha$ line, in emission in \aaa\ spectrum). The results are shown in Fig.~\ref{absD} and summarized in Table~\ref{multistar}. Given that $a$ must be null or positive, we can see that the resulting maximum distance is significantly smaller than 1~kpc, with a mean around 0.4~kpc. We emphasize here that we accounted for 40\% veiling of the disk (i.e. we multiplied the $f$ ratios in the table by a factor 0.6) before computing the distance. Any smaller fraction of the contamination will bring the object even closer to the Sun.

\section{Summary \& Conclusions}

It has been clearly shown that despite a large number of variants in distance methods, it is of prime importance to be extremely rigorous with assumptions and with the use of results made earlier and usually by others. The distance of microquasars, and in general of compact objects in our Galaxy, will benefit in a few years of the results of the satellite Gaia. Until then, we will certainly have to combine various types of information to infer good estimates to the distance of these objects. We mention here the work of \citet{Lazorenko-etal-2007} who have obtained, through astrometric measurements using VLT/FORS1, a precision of 30 {\it micro}arcseconds, similarly to what is expected for the VLTI instrument PRIMA, currently being commissioned at the ESO Cerro Paranal Observatory. This method requires a rather simple observational setup, a large field, and a large number of stars in the field. These requirements are easily fulfilled in the case of microquasars in the galactic plane like \gro, although it might again be more difficult or impossible for \aaa.

From the present work, we can conclude that:
\begin{itemize}
\item The use of the color excess measured using the sodium lines is risky. Not only one must have a spectroscopic resolution high enough, but its applicability range is limited. Moreover, one must be careful when choosing the relationship between the equivalent width and the color excess.
\item There is no proven systematic overestimation of the optical absorption as determined from X-ray data compared to that inferred from optical data. 
\item The optical data also needs to be checked for being taken truly during optical quiescence or not. This is crucial for lightcurve models.
\item The quality of the data and the uncertainties associated to it must be assessed rigorously. It is meaningless to apply a sophisticated model on data of limited quality.
\item The maximum-distance method proposed by \citet{Foellmi-etal-2006b} contains a systematic uncertainty, although its main hypothesis is sometimes implicitly used by others. 
\end{itemize}

About \gro:
\begin{itemize}
\item We have shown that although the upper limit of 3.5~kpc is a rather firm measurement, the value of 3.2 kpc has never been really measured. 
\item We have also shown that the lower limit of 3.0~kpc is questionable and actually based on a questionable assumption on the interpretation of absorption lines in the radio spectrum. Moreover, we have challenged the relevance of comparing two radio spectra in a region where there HI clouds with anomalous velocities.
\item We have estimated that $E(B-V) \sim$1.0 and $A_V \sim 3.49$. 
\item The peculiar absorption determined from X-rays towards \gro\ possibly indicate the presence of local dust close to the object. However, the X-ray flux in quiescence is also very variable.
\item The studies of the orbit and secondary star of \gro\ rely very much on the quality of the model of the disk, and on the assumed fixed parameters, often extracted from other incomplete studies. As matter of fact, no true model of the lightcurves of \gro\ combining the radial-velocity measurement and the multi-color lightcurves has been made while letting all the parameters truly vary at the same time. 
\item A new distance method using red clump giant stars applied to \gro\ however confirms a distance less than 2 kpc.
\item Although not proven, the new estimation of a  smaller distance of \gro\ strengthen the idea of its possible origin in NGC~6242. It would make \gro\ one of the closest black holes to the Sun.
\item At 1.0 kpc, $\beta = 0.28$ and there would be no misalignment between the disk and the jets, since $\theta \sim 71^{\circ}$.
\end{itemize}

As for \aaa:
\begin{itemize}
\item We have shown that the published distance and many confirmations of it are based (not always explicitely) on a single measurement of the color excess made 30 years ago on a spectrum made of 5 points, which is moreover in contradiction with an \emph{HST/STIS} spectrum.
\item We have applied the maximum-distance method of \citet{Foellmi-etal-2006b} to \aaa\ and found that it could be indeed located much closer, to a distance of $\sim$0.4~kpc.
\item The example of \aaa\ illustrate that normal distance methods based on comparing the apparent and absolute magnitudes are difficult to apply to \aaa\ since the estimation of the extinction cannot rely on relationship established in the galactic plane. For this reason also, the red clump giant star distance method could not be applied to \aaa.
\item We have also used archival images to infer an upper limit of the proper motion of \aaa, which appears quite small. Although we have found a cluster of stars 2.8 degrees away, the origin of \aaa\ remains uncertain.
\end{itemize}

Finally, to the question "What is the closest black hole to the Sun?", our answer is the black hole X-ray binary \aaa.

\section*{Acknowledgments} C.F. warmly acknowledges T.H.~Dall and E.~Depagne for advices and a critical reading of the manuscript. C.F. wants to thank G.~Nelemans for critical discussions on the details of some distance methods, E.~Moraux for help with the red clump stars, and F. Comeron for giving the necessary last bit of motivation to complete such study. C.F. is indebted to A. Burgess for proof-reading the English grammar and orthography of the manuscript. C.F. thanks the anonymous referee for corrections and pointing out an error on \aaa, and C. Sterken for insightful discussions about scientific rigor. This research has made a really extensive use of the European Southern Observatory archive, NASA's Astrophysics Data System, the arXiv astro-ph, and the Central Bureau for Astronomical Telegrams of the IAU. This research has made use of the SIMBAD database, operated at CDS, Strasbourg, France. C.F. acknowledges the Institut de Radio-Astronomie Milimetrique (IRAM) in Grenoble for keeping old volumes of $Nature$ in its library. C.F. also acknowledges the use of SuperCOSMOS Sky Surveys. Finally, C.F. acknowledges support from the Swiss National Science Foundation (grant no PA0022-115328).

\appendix
\section{Bibliographical glitches} 
\label{graph}

We mention here the few bibliographic issues that we have encountered in studying the question of the distance of \gro\ and \aaa. The \emph{Nature} paper by \citet{Bailyn-etal-1995b} quote Arnett \& Bowers in 1978 about the maximum neutron star mass, while the correct year is 1977 \citep{Arnett-Bowers-1977}. \citet{Phillips-etal-1999} quote "Bailyn et al. (1996b)" from which they use the outburst data. However the correct year is 1995. \citet{Trimble-Leonard-1995} mention the IAU Circular nb. 6063 of \citet{Inoue-etal-1994a} but attribute it for some reasons to Reynolds \& Jauncy. When looking for the X-ray fluxes of \gro\ in \citet{Garcia-etal-1998} quoted by \citet{Mirabel-etal-2002}, the authors referred to \citet{Barret-etal-1996}. However, the value of the X-ray flux is actually quoted from another paper, said to be "in press", by Zhang et al. in 1996, in the Astronomy \& Astrophysics Supplement Series. There are three publications that could match this reference: \citet{Zhang-etal-1996a} that is describing X-ray \emph{BATSE/CGRO} observations through Earth occultation and where \gro\ is not mentioned, \citet{Zhang-etal-1996b} that is about GRO~J1849--03 only and \citet{Zhang-etal-1996c}, in which only the source 4U~1608--522 is discussed. Moreover, there is no publications with Zhang as a first author in 1997 {\it and} in A\&AS. There is actually a paper in 1997 in ApJ: \citet{Zhang-etal-1997} where \emph{ASCA} and \emph{BATSE} X-ray fluxes of \gro\ are discussed, and this latter paper is certainly the correct reference. As for the color excess coming from other sources for GS 1124-684 cited by \citet{Greiner-etal-1994a}, there is a reference "West et al. (1991)". But in the reference list, we find only an IAU Circular with only "West R.M." as author, and West, R.M. Della Valle M., Jarvis B., 1991, in "Workshop on Nova Muscae 1991", Lyngby, May 1991. Interestingly, the NASA ADS system does not list any of those two references. The only seemingly reference is the IAU Circular \citet[no 5165]{West-etal-1991} but with the additional and final coauthor Pizzaro G. not mentioned in the item of the reference list of \citet{Greiner-etal-1994a}. Finally, \citet{Orosz-Bailyn-1997} refer to a submitted paper by Robinson C. et al. (1996) in the Astrophysical Journal, that cannot actually be found in NASA~ADS. We were actually not able to find any paper by C. Robinson as a first, second or third author on this subject, in 1995, 1996 or later.



\begin{thebibliography}{139}
\expandafter\ifx\csname natexlab\endcsname\relax\def\natexlab#1{#1}\fi
\expandafter\ifx\csname url\endcsname\relax
  \def\url#1{\texttt{#1}}\fi
\expandafter\ifx\csname urlprefix\endcsname\relax\def\urlprefix{URL }\fi

\bibitem[{Aannestad and Purcell(1973)}]{Aannestad-etal-1973}
Aannestad, P.~A., Purcell, E.~M., Jan 1973. Interstellar grains. \araa, 11, 309.

\bibitem[{Allen(1973)}]{Allen-1973}
Allen, C.~W., 1973. Astrophysical Quantities (Third Edition).

\bibitem[{Arnett and Bowers(1977)}]{Arnett-Bowers-1977}
Arnett, W.~D., Bowers, R.~L., 1977. A microscopic interpretation of neutron
  star structure. \apjs, 33, 415.

\bibitem[{Asai et~al.(1998)Asai, Dotani, Hoshi, Tanaka, Robinson, and
  Terada}]{Asai-etal-1998}
Asai, K., Dotani, T., Hoshi, R., Tanaka, Y., Robinson, C.~R., Terada, K., Dec
  1998. Asca observations of transient X-ray sources in quiescence. \pasj, 50,
  611.

\bibitem[{Bagnulo et~al.(2003)Bagnulo, Jehin, Ledoux, Cabanac, Melo, Gilmozzi,
  and Team}]{Bagnulo-etal-2003}
Bagnulo, S., Jehin, E., Ledoux, C., Cabanac, R., Melo, C., Gilmozzi, R., Team,
  T. E. P. S.~O., 2003. The uves paranal observatory project: A library of
  high- resolution spectra of stars across the hertzsprung-russell diagram. The
  Messenger, 114, 10--14.

\bibitem[{Bailey(1981)}]{Bailey-1981}
Bailey, J., 1981. The distances of cataclysmic variables. \mnras, 197, 31--39.

\bibitem[{Bailyn et~al.(1995{\natexlab{a}})Bailyn, Orosz, Girard, Jogee, della
  Valle, Begam, Fruchter, Gonzalez, Ianna, Layden, Martins, and
  Smith}]{Bailyn-etal-1995a}
Bailyn, C.~D., Orosz, J.~A., Girard, T.~M., Jogee, S., della Valle, M., Begam,
  M.~C., Fruchter, A.~S., Gonzalez, R., Ianna, P.~A., Layden, A.~C., Martins,
  D.~H., Smith, M., 1995{\natexlab{a}}. The optical counterpart of the
  superluminal source GRO J1655-40. \nat, 374, 701.

\bibitem[{Bailyn et~al.(1995{\natexlab{b}})Bailyn, Orosz, McClintock, and
  Remillard}]{Bailyn-etal-1995b}
Bailyn, C.~D., Orosz, J.~A., McClintock, J.~E., Remillard, R.~A.,
  1995{\natexlab{b}}. Dynamical evidence for a black hole in the eclipsing
  X-ray nova GRO~J1655-40. \nat, 378, 157.

\bibitem[{Barbon et~al.(1990)Barbon, Benetti, Rosino, Cappellaro, and
  Turatto}]{Barbon-etal-1990}
Barbon, R., Benetti, S., Rosino, L., Cappellaro, E., Turatto, M., 1990. Type ia
  supernova 1989b in ngc 3627. \aap, 237, 79--90.

\bibitem[{Barret et~al.(1996)Barret, McClintock, and
  Grindlay}]{Barret-etal-1996}
Barret, D., McClintock, J.~E., Grindlay, J.~E., Dec 1996. Luminosity
  differences between black holes and neutron stars. \apj, 473, 963.

\bibitem[{Beer and Podsiadlowski(2002)}]{Beer-Podsiadlowski-2002}
Beer, M.~E., Podsiadlowski, P., Mar 2002. The quiescent light curve and the
  evolutionary state of GRO~J1655-40. \mnras, 331, 351.

\bibitem[{Bianchini et~al.(1997)Bianchini, della Valle, Masetti, and
  Margoni}]{Bianchini-etal-1997}
Bianchini, A., della Valle, M., Masetti, N., Margoni, R., 1997. Spectroscopic
  study of GRO~J1655-40: the outburst and the decline. \aap, 321, 477--484.

\bibitem[{Bradt and McClintock(1983)}]{Bradt-etal-1983}
Bradt, H.~V.~D., McClintock, J.~E., 1983. The optical counterparts of compact
  galactic X-ray sources. \araa, 21, 13--66.

\bibitem[{Brandt et~al.(1995)Brandt, Podsiadlowski, and
  Sigurdsson}]{Brandt-etal-1995}
Brandt, W.~N., Podsiadlowski, P., Sigurdsson, S., Nov 1995. On the high space
  velocity of X-ray nova sco 1994: implications for the formation of its black
  hole. \mnras, 277, L35.

\bibitem[{Brocksopp et~al.(2006)Brocksopp, McGowan, Krimm, Godet, Roming,
  Mason, Gehrels, Still, Page, Moretti, Shrader, Campana, and
  Kennea}]{Brocksopp-etal-2006}
Brocksopp, C., McGowan, K.~E., Krimm, H., Godet, O., Roming, P., Mason, K.~O.,
  Gehrels, N., Still, M., Page, K.~L., Moretti, A., Shrader, C.~R., Campana,
  S., Kennea, J., 2006. The 2005 outburst of GRO~J1655-40: spectral evolution
  of the rise, as observed by swift. \mnras, 365, 1203--1214.

\bibitem[{Buxton and Vennes(2001)}]{Buxton-Vennes-2001}
Buxton, M.~M., Vennes, S., 2001. Atmospheric modelling of the companion star in
  GRO~J1655-40. Publications of the Astronomical Society of Australia, 18,
  91--97.

\bibitem[{Caballero-Garc\'ia et~al.(2006)Caballero-Garc\'ia, Kuulkers, Kretschmar,
  Domingo, Miller, and Mas-Hesse}]{Caballero-Garcia-etal-2006-astroph}
Caballero-Garcia, M., Kuulkers, E., Kretschmar, P., Domingo, A., Miller, J.~M.,
  Mas-Hesse, J.~M., 2006. ToO observations of GRO~J1655-40 in outburst with
  INTEGRAL. arXiv:astro-ph/0609491.

\bibitem[{Caballero-Garc\'ia et~al.(2007)Caballero-Garc\'ia, Miller, Kuulkers, D\'{i}az Trigo,
	Homan, Lewin, Kretschmar, Domingo, Mas-Hesse, Wijnands, Fabian, Fender and van der 
	Klis}]{Caballero-Garcia-etal-2007}
Caballero-Garc\'ia, M.D., Miller, J.M., Kuulkers, E., D\'iaz Trigo, M., Homan, J., 
	Lewin, W.H.G., Kretschmar, P., Domingo, A., Mas-Hesse, J.M., Wijnands, R., 
	Fabian, A.C., Fender, R.P., van der Klis, M., 2007. The High-Energy Emission of 
	GRO~J1655-40 As Revealed with INTEGRAL Spectroscopy of the 2005 Outburst. \apj, 669, 534.

\bibitem[{Cantrell et~al.(2008)Cantrell, Baylin, McClintock, 
	Orosz}]{Cantrell-etal-2008}
Cantrell, A.~G., Bailyn, C.~D., McClintock, J.~E., Orosz, J.A., 2008. Optical State Changes in the X-Ray-quiescent Black Hole A0620-00., \apj, 673, L159.

\bibitem[{Caswell et~al.(1975)Caswell, Murray, Roger, Cole, and
  Cooke}]{Caswell-etal-1975}
Caswell, J.~L., Murray, J.~D., Roger, R.~S., Cole, D.~J., Cooke, D.~J., Dec
  1975. Neutral hydrogen absorption measurements yielding kinematic distances
  for 42 continuum sources in the galactic plane. \aap, 45, 239.

\bibitem[{Chakrabarti et~al.(2008)Chakrabarti, Debnath, Nandi, 
	Pal}]{Chakrabarti-etal-2008}
Chakrabarti, S.K., Debnath, D., Nandi, A., Pal, P.S., 2008. Evolution of the 
	quasi-periodic oscillation frequency in GRO J1655--40 - Implications for 
	accretion disk dynamics. \aap, 489, L41.

\bibitem[{Cheng et~al.(1992)Cheng, Horne, Panagia, Shrader, Gilmozzi, Paresce,
  and Lund}]{Cheng-etal-1992}
Cheng, F.~H., Horne, K., Panagia, N., Shrader, C.~R., Gilmozzi, R., Paresce,
  F., Lund, N., Oct 1992. The hubble space telescope observations of X-ray nova
  muscae 1991 and its spectral evolution. \apj, 397, 664.

\bibitem[{Chevalier and Ilovaisky(1990)}]{Chevalier-Ilovaisky-1990}
Chevalier, C., Ilovaisky, S.~A., 1990. Optical studies of transient low-mass
  X-ray binaries. iii - photometry of gs 2000 + 25 during decay and quiescence.
  \aap, 238, 163--169.

\bibitem[{Clark et~al.(2003)Clark, Charles, Clarkson, and
  Coe}]{Clark-etal-2003}
Clark, J.~S., Charles, P.~A., Clarkson, W.~I., Coe, M.~J., 2003. Near ir
  spectroscopy of the X-ray binary circinus x-1. \aap, 400, 655--658.

\bibitem[{Cohen(1975)}]{Cohen-1975}
Cohen, J.~G., 1975. Optical interstellar lines in southern supergiants. \apj,
  197, 117--122.

\bibitem[{Collinder(1931)}]{Collinder-1931}
Collinder, P., 1931. On structured properties of open galactic clusters and
  their spatial distribution. Annals of the Observatory of Lund, 2, 1.

\bibitem[{Combi et~al.(2007)Combi, Bronfman, and Mirabel}]{Combi-etal-2007}
Combi, J.~A., Bronfman, L., Mirabel, I.~F., May 2007. New evidence on the
  origin of the microquasar GRO~J1655-40. \aap, 467, 597.

\bibitem[{Combi et~al.(2001)Combi, Romero, Benaglia, and
  Mirabel}]{Combi-etal-2001}
Combi, J.~A., Romero, G.~E., Benaglia, P., Mirabel, I.~F., 2001. The radio
  surroundings of the microquasar GRO~J1655-40. \aap, 370, L5--L8.

\bibitem[{Crawford and Mandwewala(1976)}]{Crawford-Mandwewala-1976}
Crawford, D.~L., Mandwewala, N., Dec 1976. Interstellar reddening relations in
  the ubv, uvby, and geneva systems. \pasp, 88, 917.

\bibitem[{Crawford et~al.(1989)Crawford, Barlow, and
  Blades}]{Crawford-etal-1989}
Crawford, I.~A., Barlow, M.~J., Blades, J.~C., Jan 1989. High-resolution
  observations of interstellar na i and ca ii absorption lines toward the
  scorpius ob1 association. Astrophysical Journal, 336, 212.

\bibitem[{dal Fiume et~al.(1999)dal Fiume, Frontera, Orlandini, Amati, del
  Sordo, Stella, Belloni, Ricci, Capalbi, and Daniele}]{dalFiume-etal-1999}
dal Fiume, D., Frontera, F., Orlandini, M., Amati, L., del Sordo, S., Stella,
  L., Belloni, T.~M., Ricci, D., Capalbi, M., Daniele, M.~R., 1999. Xte
  j1859+226. IAU Circular 7291, 2.

\bibitem[{della Valle(1994)}]{dellaValle-1994}
della Valle, M., 1994. X-ray nova in scorpius. IAU Circular 6052.

\bibitem[{della Valle and Duerbeck(1993)}]{dellaValle-Duerbeck-1993}
della Valle, M., Duerbeck, H.~W., 1993. Study of nova shells - part one -
  v1229-aquilae - nova 1970 - nebular expansion parallax and luminosity. \aap,
  275, 239.

\bibitem[{della Valle et~al.(1991)della Valle, Jarvis, and
  West}]{dellaValle-etal-1991}
della Valle, M., Jarvis, B.~J., West, R.~M., 1991. Evidence for a black hole in
  the X-ray nova muscae 1991. \nat, 353, 50--52.

\bibitem[{Dubus et~al.(1999)Dubus, Lasota, Hameury, and
  Charles}]{Dubus-etal-1999}
Dubus, G., Lasota, J.-P., Hameury, J.-M., Charles, P., Feb 1999. X-ray
  irradiation in low-mass binary systems. \mnras, 303, 139.

\bibitem[{Durant and van Kerkwijk(2006)Durant, and van Kerkwijk}]{Durant-vanKerkwijk-2006}
Durant, M., van Kerkwijk, M.~H., Distances to anomalous X-ray pulsars using red clump stars. 
	\apj, 650, 1070.

\bibitem[{Eggleton(1983)}]{Eggleton-1983}
Eggleton, P.~P., 1983. Approximations to the radii of roche lobes. \apj, 268,
  368.

\bibitem[{Esin et~al.(2000)Esin, Kuulkers, McClintock, and
  Narayan}]{Esin-etal-2000}
Esin, A.~A., Kuulkers, E., McClintock, J.~E., Narayan, R., 2000. Optical
  light curves of the black hole binaries GRS 1124-68 and A0620-00 in outburst:
  The importance of irradiation. \apj, 532, 1069.

\bibitem[{Fitzgerald(1970)}]{Fitzgerald-1970}
Fitzgerald, M.~P., 1970. The intrinsic colours of stars and two-colour
  reddening lines. \aap, 4, 234.

\bibitem[{Fitzpatrick(1999)}]{Fitzpatrick-1999}
Fitzpatrick, E.~L., Jan 1999. Correcting for the effects of interstellar
  extinction. The Publications of the Astronomical Society of the Pacific 111,
  63.

\bibitem[{Foellmi(2006)}]{Foellmi-2006e}
Foellmi, C., 2006. Proceedings of the VI Microquasar Workshop: Microquasars and Beyond, 62.

\bibitem[{Foellmi(2007)}]{Foellmi-2007c}
Foellmi, C., 2007. What really is the relativistic radio-jet distance of
  the galactic microquasar GRO~J1655-40? The Future of Photometric,
  Spectrophotometric and Polarimetric Standardization, ASP Conference Series, 364,
  597.

\bibitem[{Foellmi et~al.(2007)}]{Foellmi-etal-2007a}
Foellmi, C., Dall, T.~H, Depagne, E., 2007. On the abundances of
  GRO~J1655-40. \aap, 464, L61--L64.
  
\bibitem[{Foellmi et~al.(2006)Foellmi, Depagne, Dall, and
  Mirabel}]{Foellmi-etal-2006b}
Foellmi, C., Depagne, E., Dall, T.~H., Mirabel, I.~F., 2006. On the distance of
  GRO~J1655-40. \aap, 457, 249--255.

\bibitem[{Gallo et~al.(2006)Gallo, Fender, Miller-Jones, Merloni, Jonker,
  Heinz, Maccarone, and van~der Klis}]{Gallo-etal-2006}
Gallo, E., Fender, R.~P., Miller-Jones, J. C.~A., Merloni, A., Jonker, P.~G.,
  Heinz, S., Maccarone, T.~J., van~der Klis, M., Aug 2006. A radio-emitting
  outflow in the quiescent state of A0620-00: implications for modelling
  low-luminosity black hole binaries. \mnras, 370, 1351.

\bibitem[{Garcia et~al.(1998)Garcia, McClintock, Narayan, and
  Callanan}]{Garcia-etal-1998}
Garcia, M.~R., McClintock, J.~E., Narayan, R., Callanan, P.~J., Jan 1998. Black
  hole event horizons and X-ray nova luminosities - update. Wild Stars In The
  Old West: Proceedings of the 13th North American Workshop on Cataclysmic
  Variables and Related Objects. ASP Conference Series, 137, 506.

\bibitem[{Garcia et~al.(2001)Garcia, McClintock, Narayan, Callanan, Barret, and
  Murray}]{Garcia-etal-2001}
Garcia, M.~R., McClintock, J.~E., Narayan, R., Callanan, P.~J., Barret, D.,
  Murray, S.~S., 2001. New evidence for black hole event horizons from Chandra.
  \apjl, 553, L47--L50.

\bibitem[{Gelino et~al.(2001)Gelino, Harrison, and Orosz}]{Gelino-etal-2001b}
Gelino, D.~M., Harrison, T.~E., Orosz, J.~A., Nov 2001. A multiwavelength,
  multiepoch study of the soft X-ray transient prototype, V616 Monocerotis
  (A0620-00). \aj, 122, 2668.

\bibitem[{Gierli{\'n}ski et~al.(2001)Gierli{\'n}ski,
  Macio{\l}ek-Nied{\'z}wiecki, and Ebisawa}]{Gierlinski-etal-2001}
Gierli{\'n}ski, M., Macio{\l}ek-Nied{\'z}wiecki, A., Ebisawa, K., 2001.
  Application of a relativistic accretion disc model to X-ray spectra of LMC
  X-1 and GRO~J1655-40. \mnras, 325, 1253--1265.

\bibitem[{Gonzalez-Riestra et~al.(1991)Gonzalez-Riestra, Clavel, Cassatella,
  Krautter, and Gilmore}]{Gonzalez-Riestra-etal-1991}
Gonzalez-Riestra, R., Clavel, J., Cassatella, A., Krautter, J., Gilmore, A.~C.,
  1991. Nova in the large magellanic cloud, IAU Circular~5253.

\bibitem[{Gonz\'alez Hern\'andez et al.(2008)Gonz\'alez Hern\'andez, Rebolo, 
	Israelian}]{GonzalezHernandez-etal-2008}
Gonz\'alez Hern\'andez, J.~I., Rebolo, R., Israelian, G., 2008. The Black Hole Binary 
	Nova Scorpii 1994 (GRO J1655-40): An improved chemical analysis. \aap, 478, 203.

\bibitem[{Gottlieb and Upson(1969)}]{Gottlieb-Upson-1969}
Gottlieb, D.~M., Upson, W.~L., Aug 1969. Local interstellar reddening. \apj,
  157, 611.

\bibitem[{Gray(1992)}]{Gray-1992}
Gray, D.~F., 1992. The Observation and Analysis of Stellar Photospheres, Cambridge University Press.

\bibitem[{Greene et~al.(2001)Greene, Bailyn, and Orosz}]{Greene-etal-2001}
Greene, J., Bailyn, C.~D., Orosz, J.~A., 2001. Optical and infrared photometry
  of the microquasar GRO~J1655-40 in quiescence. \apj, 554, 1290--1297.

\bibitem[{Greiner et~al.(1994{\natexlab{a}})Greiner, Hasinger, Molendi, and
  Ebisawa}]{Greiner-etal-1994a}
Greiner, J., Hasinger, G., Molendi, S., Ebisawa, K., 1994{\natexlab{a}}.
  Near-simultaneous ROSAT and Ginga observations of the 1991 X-ray transient in
  Musca. \aap, 285, 509.

\bibitem[{Greiner et~al.(1994{\natexlab{b}})Greiner, Predehl, and
  Harmon}]{Greiner-etal-1994b}
Greiner, J., Predehl, P., Harmon, B.~A., 1994{\natexlab{b}}. ROSAT observation
  of the 4U 1543-47 outburst in 1992. AIP Conf. Series 304, 314---318.

\bibitem[{Greiner et~al.(1995)Greiner, Predehl, and Pohl}]{Greiner-etal-1995}
Greiner, J., Predehl, P., Pohl, M., 1995. ROSAT observations of GRO~J1655-40.
  \aap, 297, L67.

\bibitem[{Grindlay and Hertz(1981)}]{Grindlay-Hertz-1981}
Grindlay, J.~E., Hertz, P., 1981. Discovery of an obscured globular cluster
  associated with GX 354+0 (=4U/MXB 1728-34). \apjl, 247, L17--L21.

\bibitem[{Groenewegen et~al.(2008)Groenewegen, Udalski, and
  Bono}]{Groenewegen-etal-2008}
Groenewegen, M. A.~T., Udalski, A., Bono, G., Apr 2008. The distance to the
  galactic centre based on population II cepheids and RR Lyrae stars. \aap,
  481, 441--448.

\bibitem[{Guver et~al.(2008)Guver, Ozel, Cabrera-Lavers, and
  Wroblewski}]{Guver-etal-2008}
Guver, T., Ozel, F., Cabrera-Lavers, A., Wroblewski, P., 2008. The
  distance, mass, and radius of the neutron star in 4U 1608-52.
  ApJ submitted, arXiv:0811.3979.

\bibitem[{Harmon et~al.(1995)Harmon, Wilson, Zhang, Paciesas, Fishman,
  Hjellming, Rupen, Scott, Briggs, and Rubin}]{Harmon-etal-1995}
Harmon, B.~A., Wilson, C.~A., Zhang, S.~N., Paciesas, W.~S., Fishman, G.~J.,
  Hjellming, R.~M., Rupen, M.~P., Scott, D.~M., Briggs, M.~S., Rubin, B.~C.,
  1995. Correlations between X-ray outbursts and relativistic ejections in the
  X-ray transient GRO~J1655-40. \nat, 374, 703.

\bibitem[{Haswell et~al.(1993)Haswell, Robinson, Horne, Stiening, and
  Abbott}]{Haswell-etal-1993}
Haswell, C.~A., Robinson, E.~L., Horne, K., Stiening, R.~F., Abbott, T. M.~C.,
  1993. On the mass of the compact object in the black hole binary
  A0620-00. \apj, 411, 802.

\bibitem[{Herbig(1975)}]{Herbig-1975}
Herbig, G.~H., 1975. The diffuse interstellar bands. iv - the region
  4400-6850 a. \apj, 196, 129.

\bibitem[{Hjellming and
  Johnston(1981{\natexlab{a}})}]{Hjellming-Johnston-1981a}
Hjellming, R., Johnston, K., Mar 1981{\natexlab{a}}. Structure, strength, and
  polarization changes in radio source ss433. \nat, 290~(5802), 100--107,
  10.1038/290100a0.

\bibitem[{Hjellming and
  Johnston(1981{\natexlab{b}})}]{Hjellming-Johnston-1981b}
Hjellming, R.~M., Johnston, K.~J., 1981{\natexlab{b}}. An analysis of the
  proper motions of SS~433 radio jets. \apj, 246, L141.

\bibitem[{Hjellming and Johnston(1988)}]{Hjellming-Johnston-1988}
Hjellming, R.~M., Johnston, K.~J., 1988. Radio emission from conical jets
  associated with X-ray binaries. \apj, 328, 600.

\bibitem[{Hjellming and Rupen(1995)}]{Hjellming-Rupen-1995}
Hjellming, R.~M., Rupen, M.~P., 1995. Episodic ejection of relativistic jets by
  the X-ray transient GRO~J1655-40. \nat, 375, 464.

\bibitem[{Horne et~al.(1996)Horne, Harlaftis, Baptista, Hellier, Allan,
  Johnston, Patterson, Kemp, Haswell, and Chen}]{Horne-etal-1996}
Horne, K., Harlaftis, E.~T., Baptista, R., Hellier, C., Allan, A., Johnston,
  H., Patterson, J., Kemp, J., Haswell, C., Chen, W., 1996. GRO J1655-40. IAU
  Circular 6406, 2.

\bibitem[{Horne et~al.(1986)}]{Horne-etal-1986}
Horne, K., Wade, R.~A., Szkody, P., 1986, A dynamical model for the dwarf 
	nova AH Herculis. \mnras, 219, 791.

\bibitem[{Hynes et~al.(2002)Hynes, Haswell, Chaty, Shrader, and
  Cui}]{Hynes-etal-2002}
Hynes, R.~I., Haswell, C.~A., Chaty, S., Shrader, C.~R., Cui, W., 2002. The
  evolving accretion disc in the black hole X-ray transient XTE~J1859+226.
  \mnras, 331, 169--179.

\bibitem[{Hynes et~al.(1998)Hynes, Haswell, Shrader, Chen, Horne, Harlaftis,
  O'Brien, Hellier, and Fender}]{Hynes-etal-1998}
Hynes, R.~I., Haswell, C.~A., Shrader, C.~R., Chen, W., Horne, K., Harlaftis,
  E.~T., O'Brien, K., Hellier, C., Fender, R.~P., 1998. The 1996 outburst
  of GRO~J1655-40: the challenge of interpreting the multiwavelength spectra.
  \mnras, 300, 64.

\bibitem[{Inoue et~al.(1994)Inoue, Nagase, Ishida, Sonobe, and
  Ueda}]{Inoue-etal-1994a}
Inoue, H., Nagase, F., Ishida, M., Sonobe, T., Ueda, Y., 1994. X-ray nova in
  scorpius. IAU Circular 6063.

\bibitem[{Israelian et~al.(1999)Israelian, Rebolo, Basri, Casares and
  Mart\`{i}n}]{Israelian-etal-1999}
Israelian, G., Rebolo, R., Basri, G., Casares, J., Martin\`{i}n, E.~L., 1999. Evidence of a supernova origin for the black hole in the system GRO~J1655-40. \nat, 401, 142.

\bibitem[{Johnston et~al.(2001)Johnston, Wu, Fender, and
  Cullen}]{Johnston-etal-2001}
Johnston, H.~M., Wu, K., Fender, R.~P., Cullen, J.~G., 2001. Secular and
  orbital variability of Cir X-1 observed in optical spectra. \mnras, 328,
  1193--1199.

\bibitem[{Joinet et~al.(2008)Joinet, Kalemci, and Senziani}]{Joinet-etal-2008}
Joinet, A., Kalemci, E., Senziani, F., 2008. Hard X-ray emission of the
  microquasar GRO~J1655-40 during the rise of its 2005 outburst. \apj, 679, 655.

\bibitem[{Jonker and Nelemans(2004)}]{Jonker-Nelemans-2004}
Jonker, P.~G., Nelemans, G., Oct 2004. The distances to galactic low-mass X-ray
  binaries: consequences for black hole luminosities and kicks. \mnras, 354,
  355.

\bibitem[{Kharchenko et~al.(2005)Kharchenko, Piskunov, R{\"o}ser, Schilbach,
  and Scholz}]{Kharchenko-etal-2005}
Kharchenko, N.~V., Piskunov, A.~E., R{\"o}ser, S., Schilbach, E., Scholz,
  R.-D., 2005. Astrophysical parameters of galactic open clusters. \aap, 438,
  1163--1173.

\bibitem[{Kobayashi et~al.(2003)Kobayashi, Kubota, Nakazawa, Takahashi, and
  Makishima}]{Kobayashi-etal-2003}
Kobayashi, Y., Kubota, A., Nakazawa, K., Takahashi, T., Makishima, K., 2003.
  Observational evidence for a high-energy compton cloud in GRO~J1655-40 under
  a high accretion rate. \pasj, 55, 273--279.

\bibitem[{Kong et~al.(2002)Kong, McClintock, Garcia, Murray, and
  Barret}]{Kong-etal-2002}
Kong, A.~K.~H., McClintock, J.~E., Garcia, M.~R., Murray, S.~S., Barret, D.,
  2002. The X-ray spectra of black hole X-ray novae in quiescence as measured
  by Chandra. \apj, 570, 277--286.

\bibitem[{Koornneef(1983)}]{Koornneef-1983}
Koornneef, J., 1983. Near-infrared photometry. II - intrinsic colours and
  the absolute calibration from one to five micron. \aap, 128, 84.

\bibitem[{Kubota et~al.(2001)Kubota, Makishima, and Ebisawa}]{Kubota-etal-2001}
Kubota, A., Makishima, K., Ebisawa, K., 2001. Observational evidence for strong
  disk comptonization in GRO~J1655-40. \apjl, 560, L147--L150.

\bibitem[{Kuulkers et~al.(2000)Kuulkers, in't Zand, Cornelisse, Heise, Kong,
  Charles, Bazzano, Cocchi, Natalucci, and Ubertini}]{Kuulkers-etal-2000}
Kuulkers, E., in't Zand, J.~J.~M., Cornelisse, R., Heise, J., Kong, A.~K.~H.,
  Charles, P.~A., Bazzano, A., Cocchi, M., Natalucci, L., Ubertini, P., 2000.
  Turmoil on the accretion disk of GRO~J1655-40. \aap, 358, 993--1000.

\bibitem[{Lasota(2008)}]{Lasota-2008}
Lasota, J.-P., 2008. Adafs, accretion discs and outbursts in compact
  binaries. New Astronomy Reviews, 51, 752.

\bibitem[{Lazorenko et~al.(2007)Lazorenko, Mayor, Dominik, Pepe, Segransan, and
  Udry}]{Lazorenko-etal-2007}
Lazorenko, P.~F., Mayor, M., Dominik, M., Pepe, F., Segransan, D., Udry, S.,
  2007. High-precision astrometry on the VLT/FORS1 at time scales of few days.
  \aap, 471, 1057--1067.

\bibitem[{L{\'o}pez-Corredoira et~al.(2002)L{\'o}pez-Corredoira,
  Cabrera-Lavers, Garz{\'o}n, and Hammersley}]{LopezCorredoira-etal-2002}
L{\'o}pez-Corredoira, M., Cabrera-Lavers, A., Garz{\'o}n, F., Hammersley,
  P.~L., 2002. Old stellar galactic disc in near-plane regions according to
  2mass: Scales, cut-off, flare and warp. \aap, 394, 883.

\bibitem[{Maccarone(2002)}]{Maccarone-2002}
Maccarone, T.~J., 2002. On the misalignment of jets in microquasars. \mnras,
  336, 1371--1376.

\bibitem[{Marsh et~al.(1994)Marsh, Robinson, and Wood}]{Marsh-etal-1994}
Marsh, T.~R., Robinson, E.~L., Wood, J.~H., 1994. Spectroscopy of A0620-00 - 
	the Mass of the Black-Hole and an Image of its Accretion Disc. \mnras, 266, 137.

\bibitem[{Martin et~al.(2008)Martin, Tout, and Pringle}]{Martin-etal-2008}
Martin, R.~G., Tout, C.~A., Pringle, J.~E., 2008. Alignment time-scale of
  the microquasar GRO~J1655-40. \mnras, 387, 188.

\bibitem[{McCall(2004)}]{McCall-2004}
McCall, M.~L., 2004. On determining extinction from reddening. \aj, 128,
  2144.

\bibitem[{McClintock and Remillard(2000)}]{McClintock-Remillard-2000}
McClintock, J.~E., Remillard, R.~A., Mar 2000. HST/STIS UV spectroscopy of two
  quiescent X-ray novae: A0620-00 and Centaurus X-4. \apj, 531, 956.

\bibitem[{McClintock et~al.(2006)McClintock, Shafee, Narayan, Remillard, Davis,
  and Li}]{McClintock-etal-2006}
McClintock, J.~E., Shafee, R., Narayan, R., Remillard, R.~A., Davis, S.~W., Li,
  L.-X., 2006. The spin of the near-extreme Kerr black hole GRS~1915+105. \apj, 652, 518--539.

\bibitem[{McKay and Kesteven(1994)}]{McKay-Kesteven-1994}
McKay, D., Kesteven, M., 1994. X-ray nova in Scorpius. IAU Circular 6062.

\bibitem[{Migliari et~al.(2007)Migliari, Tomsick, Markoff, Kalemci, Bailyn, Buxton, 
	Corbel, Fender and Kaaret}]{Migliari-etal-2007}
Migliari, S., Tomsick, J.A., Markoff, S., Kalemci, E., Bailyn, C.D., Buxton, M., Corbel, 
	S., Fender, R.P., Kaaret, P., 2007. Tracing the Jet Contribution to the Mid-IR over the 
	2005 Outburst of GRO J1655-40 via Broadband Spectral Modeling. \apj, 670, 610.

\bibitem[{Mignani et~al.(2002)Mignani, Luca, Caraveo, and
  Mirabel}]{Mignani-etal-2002}
Mignani, R.~P., Luca, A.~D., Caraveo, P.~A., Mirabel, I.~F., 2002. HST
  observations rule out the association between Cir x-1 and SNR G321.9-0.3.
  \aap, 386, 487--491.

\bibitem[{Miller et~al.(2006)Miller, Raymond, Fabian, Steeghs, Homan, Reynolds,
  van~der Klis, and Wijnands}]{Miller-etal-2006}
Miller, J.~M., Raymond, J., Fabian, A.~C., Steeghs, D., Homan, J., Reynolds,
  C., van~der Klis, M., Wijnands, R., 2006. The magnetic nature of disk
  accretion onto black holes. \nat, 441, 953--955.

\bibitem[{Mirabel(2004)}]{Mirabel-2004}
Mirabel, I.~F., Oct 2004. Microquasar-AGN-GRB connections. Proceedings of the
  5th INTEGRAL Workshop on the INTEGRAL Universe (ESA SP-552), 552, 175.

\bibitem[{Mirabel et~al.(2002)Mirabel, Mignani, Rodrigues, Combi, Rodriguez,
  and Guglielmetti}]{Mirabel-etal-2002}
Mirabel, I.~F., Mignani, R., Rodrigues, I., Combi, J.~A., Rodriguez, L.~F.,
  Guglielmetti, F., 2002. The runaway black hole GRO~J1655-40. \aap, 395,
  595--599.

\bibitem[{Mirabel and Rodriguez(1994)}]{Mirabel-Rodriguez-1994}
Mirabel, I.~F., Rodriguez, L.~F., 1994. A superluminal source in the galaxy.
  \nat, 371, 46.

\bibitem[{Munari and Zwitter(1997)}]{Munari-Zwitter-1997}
Munari, U., Zwitter, T., 1997. Equivalent width of Na I and K I lines and
  reddening. \aap, 318, 269--274.

\bibitem[{Neilsen et~al.(2008)Neilsen, Steeghs, and Vrtilek}]{Neilsen-etal-2008}
Neilsen, J., Steeghs, D., Vrtilek, S.~D., 2008. The eccentric accretion disc of the 
	black hole A0620-00. \mnras, 384, 849.

\bibitem[{O'Brien et~al.(2002)O'Brien, Horne, Hynes, Chen, Haswell, and
  Still}]{OBrien-etal-2002}
O'Brien, K., Horne, K., Hynes, R.~I., Chen, W., Haswell, C.~A., Still, M.~D.,
  2002. Echoes in X-ray binaries. \mnras, 334, 426--434.

\bibitem[{Oke(1977)}]{Oke-1977}
Oke, J.~B., 1977. Further spectrophotometry of the transient X-ray source
  A0620-00. \apj, 217, 181--185.

\bibitem[{Oke and Greenstein(1977)}]{Oke-Greenstein-1977}
Oke, J.~B., Greenstein, J.~L., 1977. Spectrophotometry of the transient X-ray
  source A0620-00. \apj, 211, 872--880.

\bibitem[{Olson(1975)}]{Olson-1975}
Olson, B.~I., 1975. On the ratio of total-to-selective absorption. \pasp, 87,
  349--351.

\bibitem[{Orosz and Bailyn(1997)}]{Orosz-Bailyn-1997}
Orosz, J.~A., Bailyn, C.~D., 1997. Optical observations of GRO~J1655-40 in
  quiescence. I. A precise mass for the black hole primary. \apj, 477, 876.

\bibitem[{Orosz and Hauschildt(2000)}]{Orosz-Hauschildt-2000}
Orosz, J.~A., Hauschildt, P.~H., 2000, The use of the NextGen model atmospheres for cool giants 
in a light curve synthesis code. \aap, 364, 265--281.

\bibitem[{Orosz et~al.(1998)Orosz, Jain, Bailyn, McClintock, and
  Remillard}]{Orosz-etal-1998}
Orosz, J.~A., Jain, R.~K., Bailyn, C.~D., McClintock, J.~E., Remillard, R.~A.,
  1998. Orbital parameters for the soft X-ray transient 4U 1543-47: Evidence
  for a black hole. \apj, 499, 375.

\bibitem[{Osterbrock et~al.(1996)Osterbrock, Fulbright, Martel, Keane, Trager 
	and Basri}]{Osterbrock-etal-1996}
Osterbrock, D., Fulbright, J.~P., Martel, A.~R., Keane, M.~J., Trager, S.~C., Basri, G., 1996. Night-Sky High-Resolution Spectral Atlas of OH and O2 Emission Lines for Echelle Spectrograph Wavelength Calibration. \pasp, 108, 277.

\bibitem[{Paczy{\'n}ski(1971)}]{Paczynski-1971}
Paczy{\'n}ski, B., 1971. Evolutionary processes in close binary systems. \araa,
  9, 183.

\bibitem[{Pal and Chakrabarti(2005)}]{Pal-Chakrabarti-2005}
Pal, S., Chakrabarti, S.~K., Jun 2005. A GHz flare in a quiescent black hole
  and a determination of the mass accretion rate. Chinese Journal of Astronomy
  and Astrophysics 5, 331.

\bibitem[{Phillips et~al.(1999)Phillips, Shahbaz, and
  Podsiadlowski}]{Phillips-etal-1999}
Phillips, S.~N., Shahbaz, T., Podsiadlowski, P., 1999. The outburst radial
  velocity curve of X-ray Nova Scorpii 1994 (=GRO~J1655-40): a reduced mass
  for the black hole? \mnras, 304, 839--844.

\bibitem[{Predehl and Schmitt(1995)}]{Predehl-Schmitt-1995}
Predehl, P., Schmitt, J. H. M.~M., Jan 1995. X-raying the interstellar medium:
  ROSAT observations of dust scattering halos. \aap, 293, 889.

\bibitem[{Radhakrishnan et~al.(1972)Radhakrishnan, Goss, Murray, and
  Brooks}]{Radhakrishnan-etal-1972}
Radhakrishnan, V., Goss, W.~M., Murray, J.~D., Brooks, J.~W., Jan 1972. The
  parkes survey of 21-centimeter absorption in discrete-source spectra. III.
  21-centimeter absorption measurements on 41 galactic sources north of
  declination -48 degrees. Astrophysical Journal Supplement, 24, 49.

\bibitem[{Rees(1966)}]{Rees-1966}
Rees, M.~J., 1966. Appearance of relativistically expanding radio sources. \nat,
  211, 468.

\bibitem[{Reg{\H o}s et~al.(1998)Reg{\H o}s, Tout, and
  Wickramasinghe}]{Regos-etal-1998}
Reg{\H o}s, E., Tout, C.~A., Wickramasinghe, D., 1998. The unusual evolutionary
  state of GRO~J1655-40. \apj, 509, 362--365.

\bibitem[{Remillard et~al.(2002)Remillard, Muno, McClintock, and
  Orosz}]{Remillard-etal-2002}
Remillard, R.~A., Muno, M.~P., McClintock, J.~E., Orosz, J.~A., 2002. Evidence
  for harmonic relationships in the high-frequency quasi-periodic oscillations
  of XTE~J1550-564 and GRO~J1655-40. \apj, 580, 1030--1042.

\bibitem[{Sala et~al.(2007)Sala, Greiner, Vink, Haberl, Kendziorra, and
  Zhang}]{Sala-etal-2007}
Sala, G., Greiner, J., Vink, J., Haberl, F., Kendziorra, E., Zhang, X.~L.,
  2007. The highly ionized disk wind of GRO~J1655-40. \aap, 461, 1049--1056.

\bibitem[{S{\'a}nchez-Fern{\'a}ndez et~al.(1999)S{\'a}nchez-Fern{\'a}ndez,
  Castro-Tirado, Duerbeck, Mantegazza, Beckmann, Burwitz, Vanzi, Bianchini,
  della Valle, Piemonte, Dirsch, Hook, Yan, and
  Gim{\'e}nez}]{Sanchez-Fernandez-etal-1999}
S{\'a}nchez-Fern{\'a}ndez, C., Castro-Tirado, A.~J., Duerbeck, H.~W.,
  Mantegazza, L., Beckmann, V., Burwitz, V., Vanzi, L., Bianchini, A., della
  Valle, M., Piemonte, A., Dirsch, B., Hook, I., Yan, L., Gim{\'e}nez, A.,
  1999. Optical observations of the black hole candidate XTE~J1550-564 during
  the september/october 1998 outburst. \aap, 348, L9--L12.

\bibitem[{Schmidt(1965)}]{Schmidt-1965}
Schmidt, M., Jan 1965. Rotation parameters and distribution of mass in the
  galaxy. Galactic structure. Edited by Adriaan Blaauw and Maarten Schmidt, University of Chicago Press, 513.

\bibitem[{Shahbaz(2003)}]{Shahbaz-2003}
Shahbaz, T. 2003. Determining the spectroscopic mass ratio in interacting binaries: 
application to X-Ray Nova Sco 1994, \mnras, 339, 1031

\bibitem[{Shahbaz et~al.(2004)Shahbaz, Hynes, Charles, Zurita, Casares,
  Haswell, Araujo-Betancor, and Powell}]{Shahbaz-etal-2004}
Shahbaz, T., Hynes, R.~I., Charles, P.~A., Zurita, C., Casares, J., Haswell,
  C.~A., Araujo-Betancor, S., Powell, C., Oct 2004. Optical spectroscopy of
  flares from the black hole X-ray transient A0620-00 in quiescence. \mnras,
  354, 31.

\bibitem[{Shahbaz et~al.(1994)Shahbaz, Naylor, and Charles}]{Shahbaz-etal-1994}
Shahbaz, T., Naylor, T., Charles, P.~A., Jun 1994. The mass of the black hole
  in A0620-00. \mnras, 268, 756.

\bibitem[{Shahbaz et~al.(1999)Shahbaz, van~der Hooft, Casares, Charles, and van
  Paradijs}]{Shahbaz-etal-1999}
Shahbaz, T., van~der Hooft, F., Casares, J., Charles, P.~A., van Paradijs, J.,
  1999. The mass of X-ray nova Scorpii1994 (=GRO~J1655-40). \mnras, 306, 89--94.

\bibitem[{Shaposhnikov and Titarchuk(2009)Shaposhnikov, and 
	Titarchuk}]{Shaposhnikov-Titarchuk-2009}
Shaposhnikov, N., Titarchuk, L., 2009	. Determination of Black Hole Masses in 
	Galactic Black Hole Binaries using Scaling of Spectral and Variability 
	Characteristics. arXiv:09022852v1.

\bibitem[{Shaver et~al.(1982)Shaver, Radhakrishnan, Anantharamaiah, Retallack,
  Wamsteker, and Danks}]{Shaver-etal-1982}
Shaver, P.~A., Radhakrishnan, V., Anantharamaiah, K.~R., Retallack, D.~S.,
  Wamsteker, W., Danks, A.~C., 1982. Anomalous motions of H I clouds. \aap,
  106, 105.

\bibitem[{Soria et~al.(2000)Soria, Wu, and Hunstead}]{Soria-etal-2000}
Soria, R., Wu, K., Hunstead, R.~W., 2000. Optical spectroscopy of GRO~J1655-40.
  \apj, 539, 445--462.

\bibitem[{Spitzer(1948)}]{Spitzer-1948}
Spitzer, L., Sep 1948. The distribution of interstellar sodium. \apj, 108, 276.

\bibitem[{Stebbins et~al.(1940)Stebbins, Huffer, and
  Whitford}]{Stebbins-etal-1940}
Stebbins, J., Huffer, C.~M., Whitford, A.~E., Jan 1940. The colors of 1332 B
  stars. \apj, 91, 20.

\bibitem[{Stevens et~al.(2003)Stevens, Hannikainen, Wu, Hunstead, and
  McKay}]{Stevens-etal-2003}
Stevens, J.~A., Hannikainen, D.~C., Wu, K., Hunstead, R.~W., McKay, D.~J.,
  2003. The radio flaring behaviour of GRO~J1655-40: an analogy with
  extragalactic radio sources? \mnras, 342, 623--628.

\bibitem[{Takahashi et~al.(2008)Takahashi, Fukazawa, Mizuno, Hirasawa,
  Kitamoto, Sudoh, Ogita, Kubota, Makishima, Itoh, Parmar, Ebisawa, Naik,
  Dotani, Kokubun, Ohnuki, Takahashi, Yaqoob, Angelini, Ueda, Yamaoka, Kotani,
  Kawai, Namiki, Kohmura, and Negoro}]{Takahashi-etal-2008}
Takahashi, H., Fukazawa, Y., Mizuno, T., Hirasawa, A., Kitamoto, S., Sudoh, K.,
  Ogita, T., Kubota, A., Makishima, K., Itoh, T., Parmar, A.~N., Ebisawa, K.,
  Naik, S., Dotani, T., Kokubun, M., Ohnuki, K., Takahashi, T., Yaqoob, T.,
  Angelini, L., Ueda, Y., Yamaoka, K., Kotani, T., Kawai, N., Namiki, M.,
  Kohmura, T., Negoro, H., 2008. Low/hard state spectra of GRO~J1655-40
  observed with Suzaku. Publications of the Astronomical Society of Japan, 60,
  69.

\bibitem[{Tingay et~al.(1995)Tingay, Jauncey, Preston, Reynolds, Meier, Murphy,
  Tzioumis, McKay, Kesteven, Lovell, Campbell-Wilson, Ellingsen, Gough,
  Hunstead, Jones, McCulloch, Migenes, Quick, Sinclair, and
  Smith}]{Tingay-etal-1995}
Tingay, S.~J., Jauncey, D.~L., Preston, R.~A., Reynolds, J.~E., Meier, D.~L.,
  Murphy, D.~W., Tzioumis, A.~K., McKay, D.~J., Kesteven, M.~J., Lovell,
  J.~E.~J., Campbell-Wilson, D., Ellingsen, S.~P., Gough, R., Hunstead, R.~W.,
  Jones, D.~L., McCulloch, P.~M., Migenes, V., Quick, J., Sinclair, M.~W.,
  Smith, D., 1995. Relativistic motion in a nearby bright X-ray source. \nat,
  374, 141.

\bibitem[{Trimble and Leonard(1995)}]{Trimble-Leonard-1995}
Trimble, V., Leonard, P.~J.~T., 1995. Astrophysics in 1994. \pasp, 107, 1--21.

\bibitem[{Tsunemi et~al.(1989)Tsunemi, Kitamoto, Okamura, and
  Roussel-Dupre}]{Tsunemi-etal-1989}
Tsunemi, H., Kitamoto, S., Okamura, S., Roussel-Dupre, D., 1989. Discovery of a
  bright X-ray nova, GS~2000+25. \apjl, 337, L81--L84.

\bibitem[{Ueda et~al.(1998)Ueda, Inoue, Tanaka, Ebisawa, Nagase, Kotani, and
  Gehrels}]{Ueda-etal-1998}
Ueda, Y., Inoue, H., Tanaka, Y., Ebisawa, K., Nagase, F., Kotani, T., Gehrels,
  N., Jan 1998. Detection of absorption-line features in the X-ray spectra of
  the galactic superluminal source GRO~J1655-40. \apj, 492, 782.

\bibitem[{van~der Hooft {et~al.}(1997)van~der Hooft, Groot, Shahbaz,
  Augusteijn, Casares, Dieters, Greenhill, Hill, Scheers, Naber, de~Jong,
  Charles, \& van Paradijs}]{vanderHooft-etal-1997}
van~der Hooft, F., Groot, P.~J., Shahbaz, T., {et~al.} 1997. The black hole transient Nova Scorpii 1994 (=GRO J1655-40): orbital ephemeris and optical light curve, \mnras, 286, L43

\bibitem[{van~der Hooft et~al.(1998)van~der Hooft, Heemskerk, Alberts, and van
  Paradijs}]{vanderHooft-etal-1998}
van~der Hooft, F., Heemskerk, M.~H.~M., Alberts, F., van Paradijs, J., 1998.
  The quiescence optical light curve of Nova Scorpii 1994 (=GRO~J1655-40). \aap,
  329, 538--550.

\bibitem[{van~der Woerd et~al.(1989)van~der Woerd, White, and
  Kahn}]{vanderWoerd-etal-1989}
van~der Woerd, H., White, N.~E., Kahn, S.~M., 1989. X-ray spectroscopy of the
  ultrasoft transient 4U~1543-47. \apj, 344, 320--324.

\bibitem[{Vrtilek et~al.(1991)Vrtilek, McClintock, Seward, Kahn, and
  Wargelin}]{Vrtilek-etal-1991}
Vrtilek, S.~D., McClintock, J.~E., Seward, F.~D., Kahn, S.~M., Wargelin, B.~J.,
  1991. The Einstein objective grating spectrometer survey of galactic binary
  X-ray sources. \apjs, 76, 1127--1167.

\bibitem[{Wagner et~al.(1994)Wagner, Starrfield, Hjellming, Howell, and
  Kreidl}]{Wagner-etal-1994}
Wagner, R.~M., Starrfield, S.~G., Hjellming, R.~M., Howell, S.~B., Kreidl,
  T.~J., 1994. ROSAT observations of the black hole candidate V404 Cygni in
  quiescence. \apjl, 429, L25--L28.

\bibitem[{Wampler(1966)}]{Wampler-1966}
Wampler, E.~J., Jun 1966. Scanner observations of $\lambda$~4430. \apj, 144, 921.

\bibitem[{Welsh et~al.(1990)Welsh, Vedder, and Vallerga}]{Welsh-etal-1990}
Welsh, B.~Y., Vedder, P.~W., Vallerga, J.~V., 1990. High-resolution sodium
  absorption-line observations of the local interstellar medium. \apj, 358, 473.

\bibitem[{West et~al.(1991)West, della Valle, Jarvis, and
  Pizarro}]{West-etal-1991}
West, R.~M., della Valle, M., Jarvis, B., Pizarro, G., 1991. Nova Muscae 1991
  (X-ray transient in Musca). IAU Circular~5165.

\bibitem[{Willems et~al.(2005)Willems, Henninger, Levin, Ivanova, Kalogera,
  McGhee, Timmes, and Fryer}]{Willems-etal-2005}
Willems, B., Henninger, M., Levin, T., Ivanova, N., Kalogera, V., McGhee, K.,
  Timmes, F.~X., Fryer, C.~L., 2005. Understanding compact object formation and
  natal kicks. I. Calculation methods and the case of GRO~J1655-40. \apj, 625,
  324--346.

\bibitem[{Wu et~al.(1976)Wu, Aalders, van Duinen, Kester, and
  Wesselius}]{Wu-etal-1976}
Wu, C.-C., Aalders, J.~W.~G., van Duinen, R.~J., Kester, D., Wesselius, P.~R.,
  1976. A study of the transient X-ray source A0620-00. \aap, 50, 445--449.

\bibitem[{Wu et~al.(1983)Wu, Panek, Holm, Schmitz, and Swank}]{Wu-etal-1983}
Wu, C.-C., Panek, R.~J., Holm, A.~V., Schmitz, M., Swank, J.~H., 1983.
  Ultraviolet observations of the transient X-ray sources A0535+26 and
  A0620-00. \pasp, 95, 391--397.

\bibitem[{Xiang et~al.(2007)Xiang, Lee, and Nowak}]{Xiang-etal-2007}
Xiang, J., Lee, J.~C., Nowak, M.~A., 2007. Using the X-ray dust scattering
  halo of 4U 1624-490 to determine distance and dust distributions. \apj, 660,
  1309.

\bibitem[{Yamaoka et~al.(2001)Yamaoka, Ueda, Inoue, Nagase, Ebisawa, Kotani,
  Tanaka, and Zhang}]{Yamaoka-etal-2001}
Yamaoka, K., Ueda, Y., Inoue, H., Nagase, F., Ebisawa, K., Kotani, T., Tanaka,
  Y., Zhang, S.~N., 2001. Asca observation of the superluminal jet source GRO~J1655-40 in the 1997 outburst. \pasj, 53, 179--188.

\bibitem[{York(1971)}]{York-1971}
York, D.~G., May 1971. Structure in the interstellar-extinction curve.
  \apj, 166, 65.

\bibitem[{Zhang et~al.(1997)Zhang, Ebisawa, Sunyaev, Ueda, Harmon, Sazonov,
  Fishman, Inoue, Paciesas, and Takahashi}]{Zhang-etal-1997}
Zhang, S.~N., Ebisawa, K., Sunyaev, R., Ueda, Y., Harmon, B.~A., Sazonov, S.,
  Fishman, G.~J., Inoue, H., Paciesas, W.~S., Takahashi, T., 1997. Broadband
  high-energy observations of the superluminal jet source GRO~J1655-40 during
  an outburst. \apj, 479, 381.

\bibitem[{Zhang et~al.(1996{\natexlab{a}})Zhang, Harmon, Paciesas, and
  Fishman}]{Zhang-etal-1996a}
Zhang, S.~N., Harmon, B.~A., Paciesas, W.~S., Fishman, G.~J.,
  1996{\natexlab{a}}. Deep search for celestial hard X-ray emission by Earth
  occultation with BATSE/CGRO. \aaps 120, C137+.

\bibitem[{Zhang et~al.(1996{\natexlab{b}})Zhang, Harmon, Paciesas, Fishman,
  Finger, Robinson, Rubin, Grindlay, Barret, Tavani, Kaaret, Bloser, and
  Ford}]{Zhang-etal-1996b}
Zhang, S.~N., Harmon, B.~A., Paciesas, W.~S., Fishman, G.~J., Finger, M.~H.,
  Robinson, C.~R., Rubin, B.~C., Grindlay, J.~E., Barret, D., Tavani, M.,
  Kaaret, P., Bloser, P., Ford, E., 1996{\natexlab{b}}. Periodic transient hard
  X-ray emission from GRO~1849-03. \aaps 120, C227.

\bibitem[{Zhang et~al.(1996{\natexlab{c}})Zhang, Harmon, Paciesas, Fishman,
  Grindlay, Barret, Tavani, Kaaret, Bloser, Ford, and
  Titarchuk}]{Zhang-etal-1996c}
Zhang, S.~N., Harmon, B.~A., Paciesas, W.~S., Fishman, G.~J., Grindlay, J.~E.,
  Barret, D., Tavani, M., Kaaret, P., Bloser, P., Ford, E., Titarchuk, L.,
  1996{\natexlab{c}}. Low state hard X-ray outburst from the X-ray burster 4U~1608-522 observed by BATSE/CGRO. \aaps 120, C279.

\bibitem[{Zhang et~al.(1994)Zhang, Wilson, Harmon, Fishman, Wilson, Paciesas,
  Scott, and Rubin}]{Zhang-etal-1994}
Zhang, S.~N., Wilson, C.~A., Harmon, B.~A., Fishman, G.~J., Wilson, R.~B.,
  Paciesas, W.~S., Scott, M., Rubin, B.~C., 1994. X-ray nova in scorpius. IAU
  Circular 6046, 1.

\end{thebibliography}
\end{document}